\documentclass{PoS}
\pdfpageattr {/Group << /S /Transparency /I true /CS /DeviceRGB>>}
\usepackage{amsmath}
\usepackage{multicol,lipsum}
\usepackage{xspace}
\usepackage{cite}
\usepackage{bm}
\usepackage{caption}
\usepackage{subcaption}
\newcommand{\signal}{\ensuremath{S(1000)}\xspace}
\newcommand{\xmax}{\ensuremath{X_\mathrm{max}}\xspace}
\newcommand{\xmaxmed}{\ensuremath{\left< X_\mathrm{max} \right>}\xspace}
\newcommand{\sigXmax}{\ensuremath{\sigma(X_\mathrm{max})}\xspace}
\newcommand{\xmaxsd}{\ensuremath{X_\mathrm{max}^{SD}}\xspace}

\newcommand{\gcmd}{\ensuremath{\mathrm{g~cm^{-2}/decade}}\xspace}

\newcommand{\lnA}{\ensuremath{\ln A}\xspace}

\newcommand{\erange}[2]{\ensuremath{10^{#1}-10^{#2}~{\rm eV}}\xspace}
\newcommand{\exprange}{\ensuremath{\mathrm {km^{2}~sr~yr}}\xspace}
\newcommand{\frange}{\ensuremath{\mathrm {GeV~cm^{-2}~sr^{-1}~s^{-1}}}\xspace}
\title{Highlights from the Pierre Auger Observatory}

\ShortTitle{Auger Highlights}

\author{\speaker{Antonella Castellina}$^a$ for the Pierre Auger Collaboration$^b$\\
\llap{$^a$} Osservatorio Astrofisico di Torino (INAF) and INFN, Sezione di Torino, Italy\\
\llap{$^b$}Observatorio Pierre Auger, Av.\ San Mart\'in Norte 304, 5613 Malarg\"ue, Argentina\\
E-mail: \href{mailto:auger_spokespersons@fnal.gov}{\rm auger\_spokespersons@fnal.gov}\\
Full author list: \href{http://www.auger.org/archive/authors_icrc_2019.html}{\rm http://www.auger.org/archive/authors\_icrc\_2019.html}}

\abstract{
The Pierre Auger Observatory has collected data for more than 15 years, accumulating the world's 
largest exposure to ultrahigh energy cosmic rays (UHECRs).  The energy spectrum of the UHECRs 
has been measured up to $10^{20.2}$ eV, and their mass composition has been studied exploiting
both the fluorescence and surface detectors of the Observatory.  The analysis of the UHECRs arrival 
directions allowed us to measure their large-scale anisotropies over more than three decades in energy. 
The searches for intermediate-scale anisotropies are being performed with an exposure now reaching 
approximately $100,000$ \exprange.
For the first time, the fluctuations in the number of muons in inclined air showers above $4  \times 10^{18}$ 
eV have been obtained, while more information in the energy range  \erange{17.5}{18.3} comes from the 
direct measure of muons with the underground scintillators of the Observatory. Moreover, we discuss 
the limits obtained on the fluxes of neutral particles and the results of the searches for UHE neutrinos 
from transient astrophysical sources.
The new perspectives opened by the current results call for an upgrade of the Observatory, whose 
main aim is the collection of new information about the primary mass of the highest energy cosmic rays 
to understand the origin of the observed flux suppression. The progress of the upgrade of the Observatory, 
AugerPrime, is presented and discussed.
}

\FullConference{36th International Cosmic Ray Conference --- ICRC2019\\
                24 July -- 1 August, 2019\\
                Madison, Wisconsin, USA}

\begin{document}
%%% FIG 1 
\begin{figure}[b]
  \includegraphics[width=.5\linewidth]{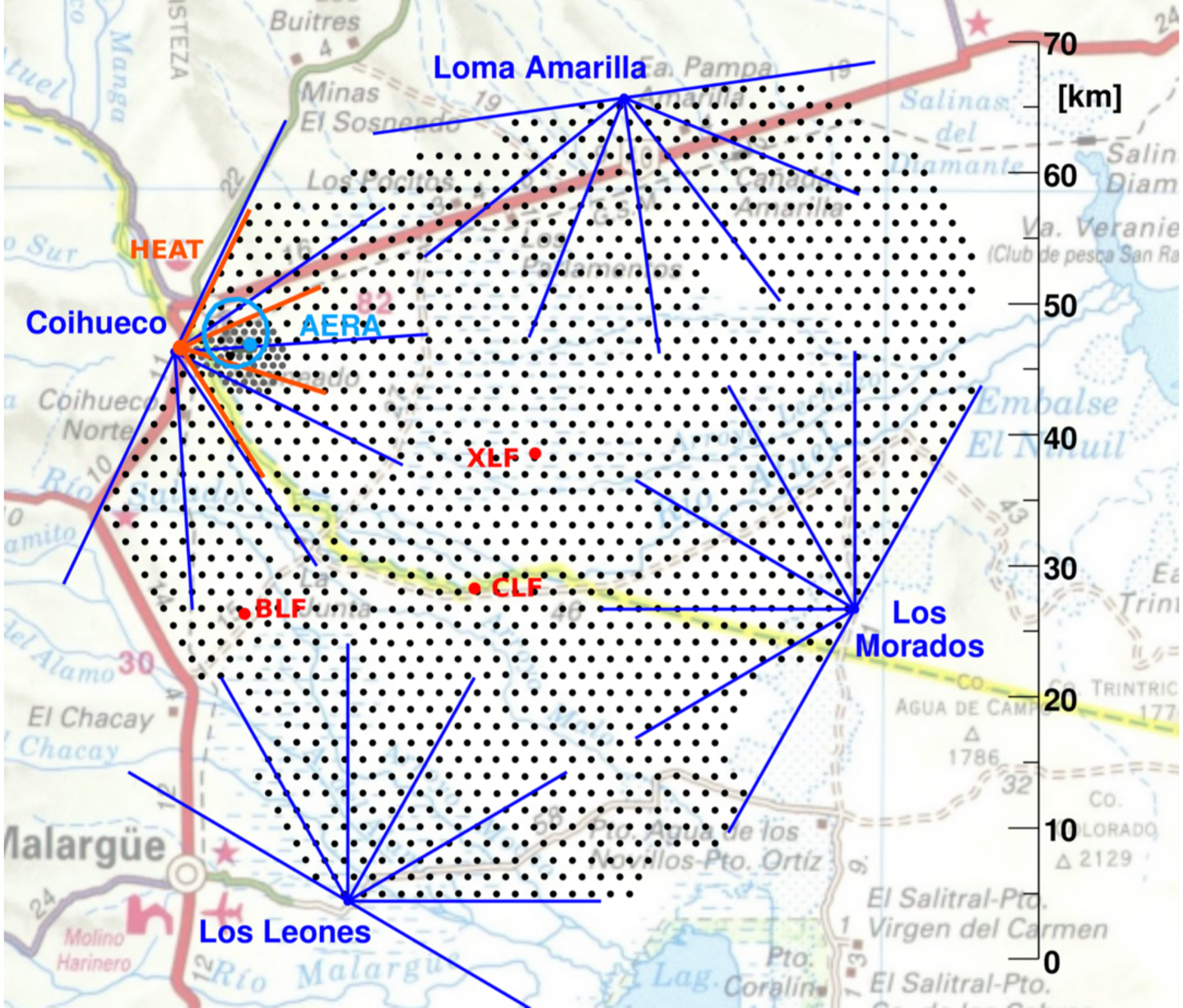}
  \includegraphics[width=.5\linewidth]{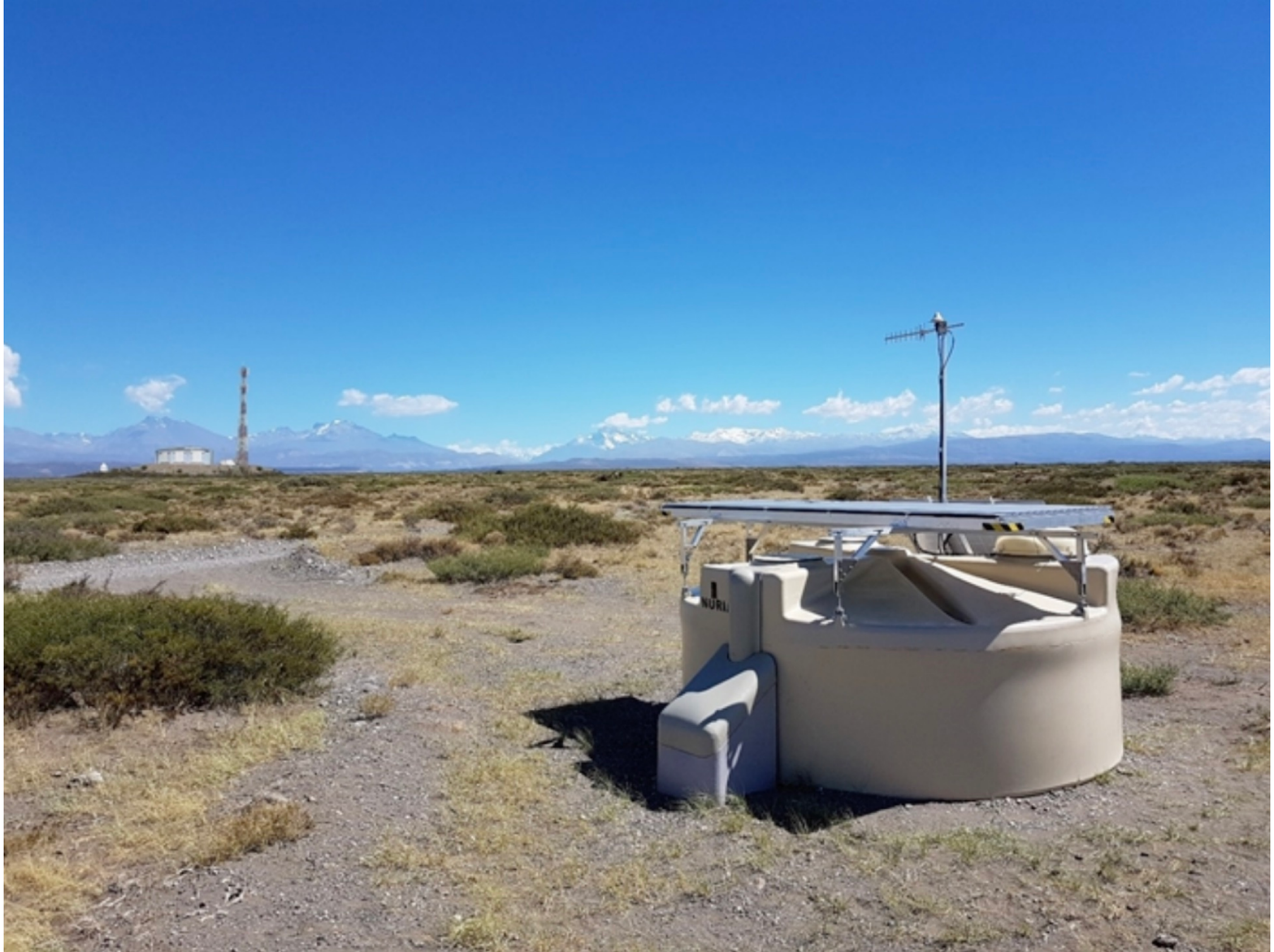}
\caption{Left: layout of the Pierre Auger Observatory, showing the WCD (black dots) and the azimuthal 
field of views of the 27 FD (blue and red lines). The two laser facilities (CLF and XLF) for the monitoring 
of the aerosol content in the atmosphere are indicated by the red dots, the AERA detectors in cyan. 
Right: photograph of one of the (upgraded) WCD stations, with a FD site visible in the background.}
  \label{fig:photo}
\end{figure}
\section{Introduction}
The Pierre Auger Observatory was built more than 15 years ago thanks to the contribution of more than
400 scientists from 17 different countries and to the essential help of a team of on-site physicists and 
technicians, with the primary scientific aim of studying cosmic rays in the ultrahigh energy range above 
$\sim 10^{17}$ eV, to understand their nature and to find their sources. 
Its layout  is shown in the left panel of Fig.\ref{fig:photo}.
The surface detector (SD1500) is formed by 1600  water Cherenkov stations (WCD) deployed in a 1.5 
km grid over a total area of about 3000 km$^2$, and is used to measure the density and the arrival time 
distribution of the shower particles at the ground. 
This area is overlooked by four fluorescence detectors (FD), each consisting of six telescopes with elevation 
angle from 1$^{\circ}$ to 30$^{\circ}$, for the measurement of the energy deposit of the showers in the atmosphere.
A photograph of one of the SD stations is shown in the right panel of Fig.\ref{fig:photo}.
In the North-Western part of the experimental site, 61 WCD (SD750) are deployed in a denser array 
with a detector spacing of 750 m, together with three high-elevation fluorescence telescopes (HEAT), with 
elevation angle from 30$^{\circ}$ to 60$^{\circ}$. This region is exploited to widen the explorable energy region 
down to about $10^{16.5}$ eV.  
The Auger Engineering Radio Array (AERA) complements the measurements of low energy showers by 
detecting their radio emission, using more than 150 radio antennas.
Different atmospheric monitoring devices are employed to monitoring the atmosphere.  
All details about the Observatory can be found in \cite{nim}. 

The rich harvest of the Pierre Auger Collaboration covers different and complementary fields of research. 
The quest for the origin of UHE cosmic rays relies on the measurement of the energy spectrum and 
mass composition of the primaries, on multi-messenger studies and on extensive searches for anisotropy 
at both large and intermediate angular scale; our most updated results are discussed in Sec.\ref{sec:astrop}.  
The properties of multiparticle production are explored thanks to the collection of different observables linked to
the muonic and electromagnetic components of the showers; the outcome is presented in Sec.\ref{sec:hep}.
Our current efforts in upgrading the Observatory are summarized in Sec.\ref{sec:upgrade}.
Due to the limited number of pages, only selected results are presented here; we refer the reader to 
\cite{arr1,cg1,cg2,found2,found4,joint1,joint4,out1} for  further interesting studies from the Pierre Auger 
Observatory presented at this conference.  All contributions are available within one volume in \cite{arx}.

\section{Astrophysics of UHE cosmic rays}
\label{sec:astrop}

\subsection{Energy spectrum}

A huge exposure of about 60,000 \exprange allows us to measure the cosmic ray energy spectrum, 
exploiting more than 215,000 events collected by the SD1500 array with energies above $2.5 \times 10^{18}$ 
eV and arriving at zenith angles less than $60^{\circ}$ \cite{spec1}.\\
The reconstruction of these events is described in \cite{found7}.
The signals from through-going atmospheric muons, collected every minute, are used to convert the 
particle densities in each SD station into a signal in units of Vertical Equivalent Muons (VEM), providing a 
common reference to all stations.
The accuracy in the evaluation of the arrival direction of the shower depends on the start times of the 
individual stations; it is derived by means of Monte Carlo simulations based on the EPOS-LHC model, 
and at the highest 
energies it results to be better than $1^{\circ}$ for all simulated angles and masses. \\
The energy estimator of the SD is chosen to be the signal evaluated at 1 km from the core, where the
fluctuations in the value of \signal due to uncertainties in the lateral distribution function are minimized. 
The uncertainty on \signal decreases from 15\% to 5\% for shower sizes increasing from 10 to 250 VEM.  
By applying the Constant Intensity Cut (CIC) method, \signal is corrected for the attenuation of
showers when propagating through different amounts of atmosphere, normalizing it to the average arrival 
direction of $38^{\circ}$ ($S_{38}$).
We employ a new parametrization of the CIC which takes into account the energy dependence of the shower 
attenuation \cite{spec1}, due to changes in composition with increasing energy or, for fixed primary mass, 
to the evolution of the electromagnetic fraction with respect to the total signal.

\begin{figure}[t]
  \includegraphics[width=.49\linewidth]{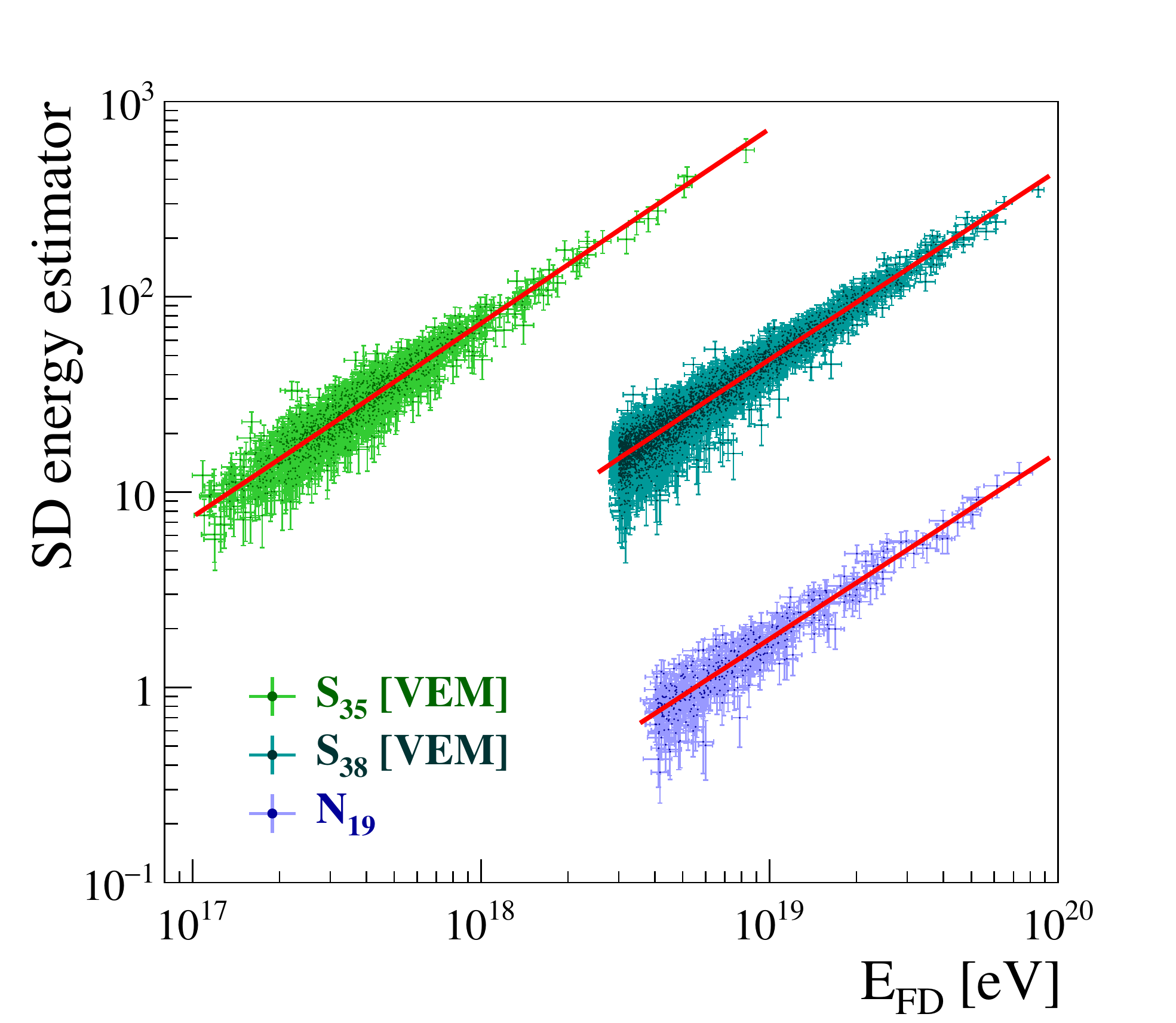}
  \includegraphics[width=.51\linewidth]{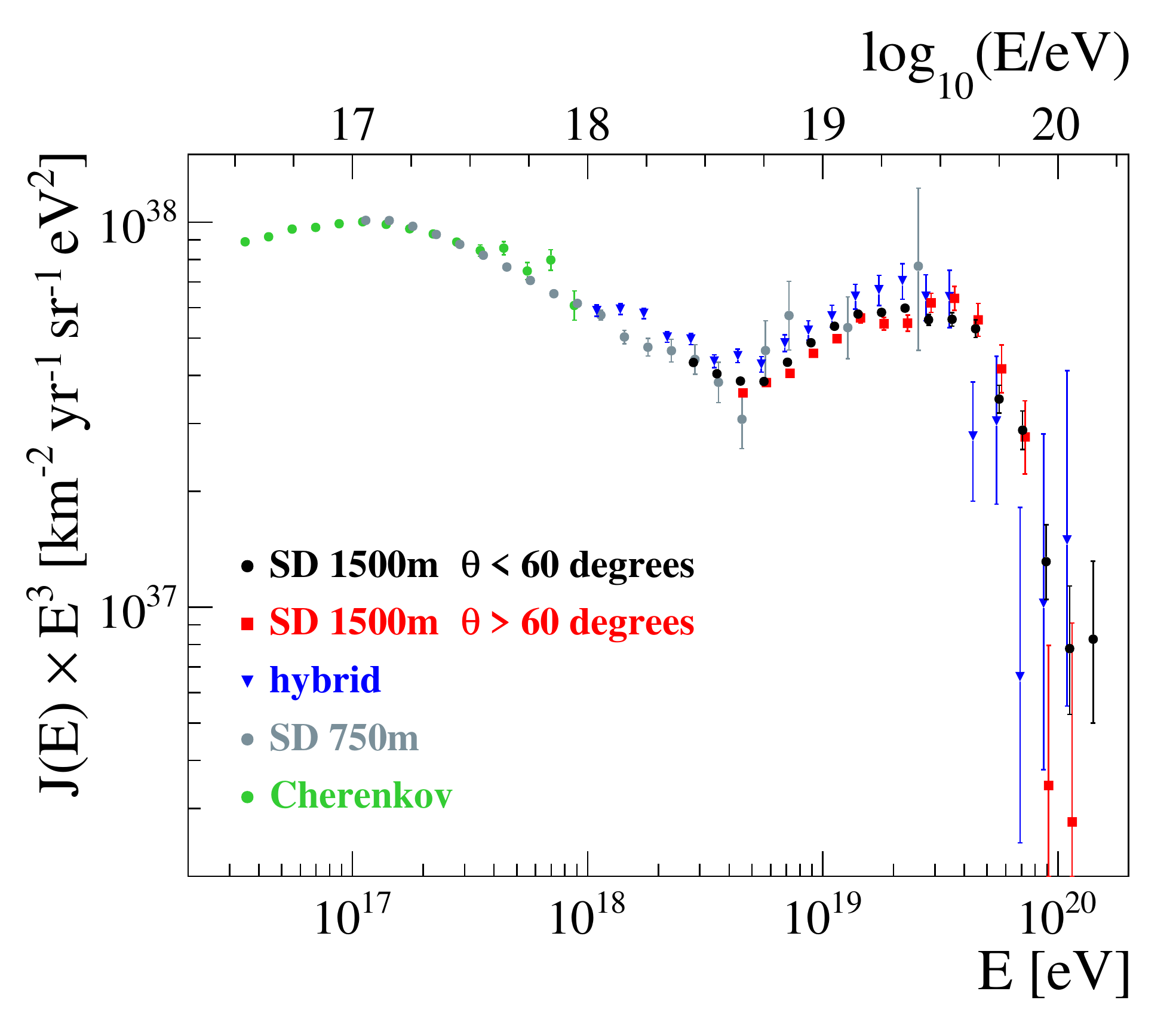}
  \caption{Left: Correlation between the FD energy and the SD energy estimators as measured using the 
  different data sets.  Right:  Energy spectra measured at the Pierre Auger Observatory.}
  \label{fig:spettri}
\end{figure}

Using the FD technique, we can measure the nearly calorimetric energy deposited in the atmosphere by 
the electromagnetic component of the EAS. Exploiting the correlation of the SD energy estimator $S_{38}$ 
with the FD energy in hybrid events, i.e. those detected by both the SD and FD,  we obtain the energy 
scale of the Observatory and a calibration of all SD events in a model-independent way. This correlation 
is shown in the left panel of Fig.\ref{fig:spettri} for the SD1500 vertical sample, together with those measured 
with the SD750 and the SD1500 horizontal data sets.  For the latter, a completely different reconstruction is 
performed, yielding to the energy estimator $N_{19}$, the relative muon content with respect to simulated 
proton showers at 10 EeV \cite{horrec}.\\
At this conference, we presented a review of the energy scale of the Observatory, discussing the FD energy 
resolution and further checks on the systematic uncertainties  \cite{found6}.
Real-time measurements of the atmospheric conditions are mandatory for the correct evaluation of the 
fluorescence light emitted by the showers. The main contribution to their systematic uncertainties comes from 
the determination of the aerosol content; in particular, we showed the importance of using hourly aerosols 
quantification to avoid a significant worsening  of the resolution on both the energy and the \xmax 
determination \cite{det1}. The reconstruction of the shower profile has been improved by using a new function, 
where the shape parameters are free from correlations, in this way allowing better-behaved fits \cite{RLfit}.
The invisible energy, i.e. that carried by particles that do not dissipate all their energy in the atmosphere, is
derived in a fully data-driven approach \cite{einv}, as such free from the systematic uncertainties affecting 
the hadronic interaction models used in the simulations. Its current parametrization is based on the horizontal 
events (muon dominated)  and adds up to a fraction of 20 to 12\% of the total energy ranging from $10^{17}$ 
to $10^{20}$ eV, with systematic uncertainties of less than 2\%. 
The FD energy resolution amounts to about 8\%, adding the contributions coming from the atmosphere, the 
detector and reconstruction and the invisible energy. We confirm the overall systematic uncertainty of 14\% 
on the energy scale \cite{esys}.

The vertical energy spectrum is shown in the right panel of Fig.\ref{fig:spettri} together with those from 
the other independent and complementary data sets. 
% calib + spettri
\begin{figure}
\begin{minipage}[c]{7cm}
\includegraphics[width=6.5cm, height=5.cm]{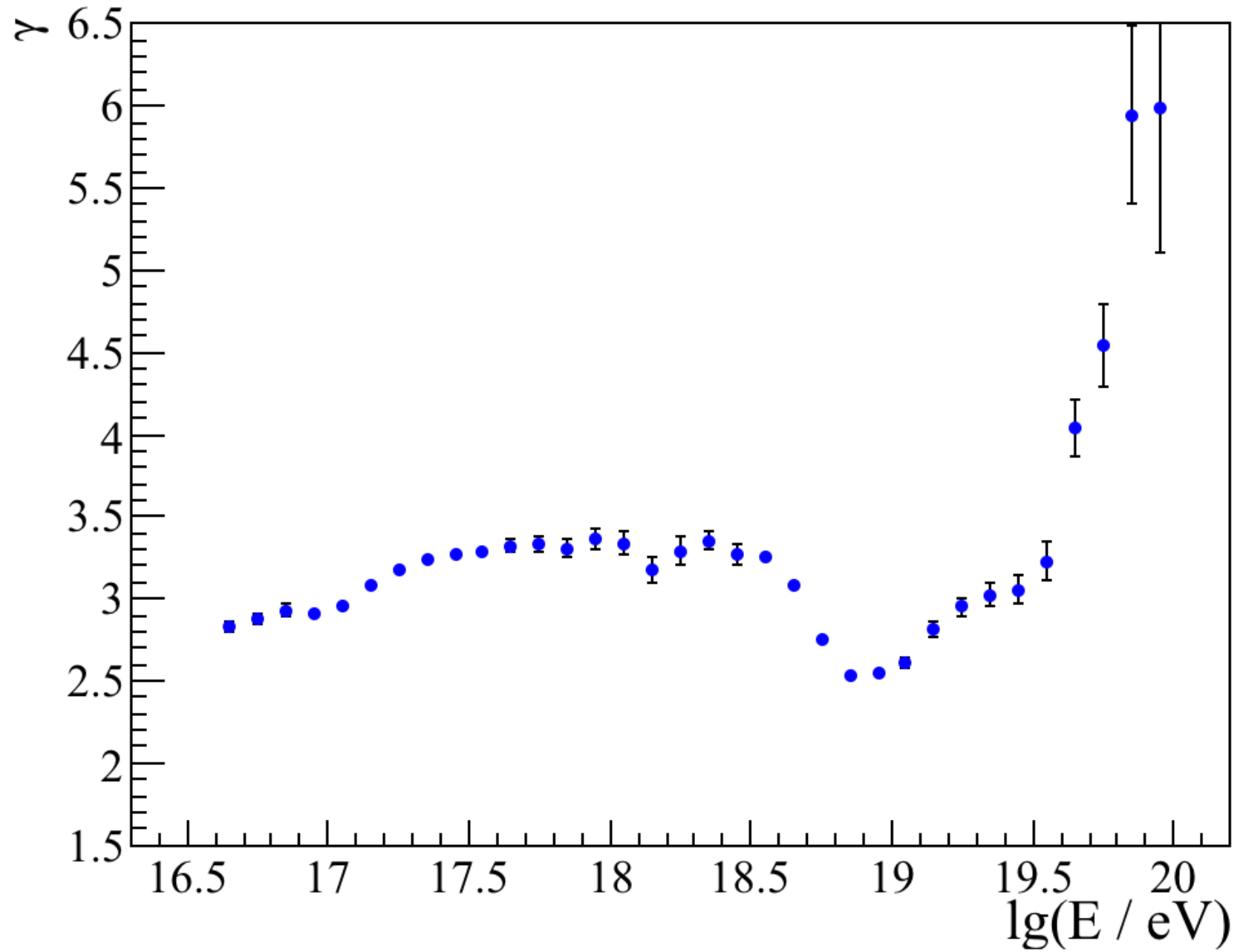}
\end{minipage}
\begin{minipage}[c]{9cm}
\includegraphics[width=8.5cm, height=7.0cm]{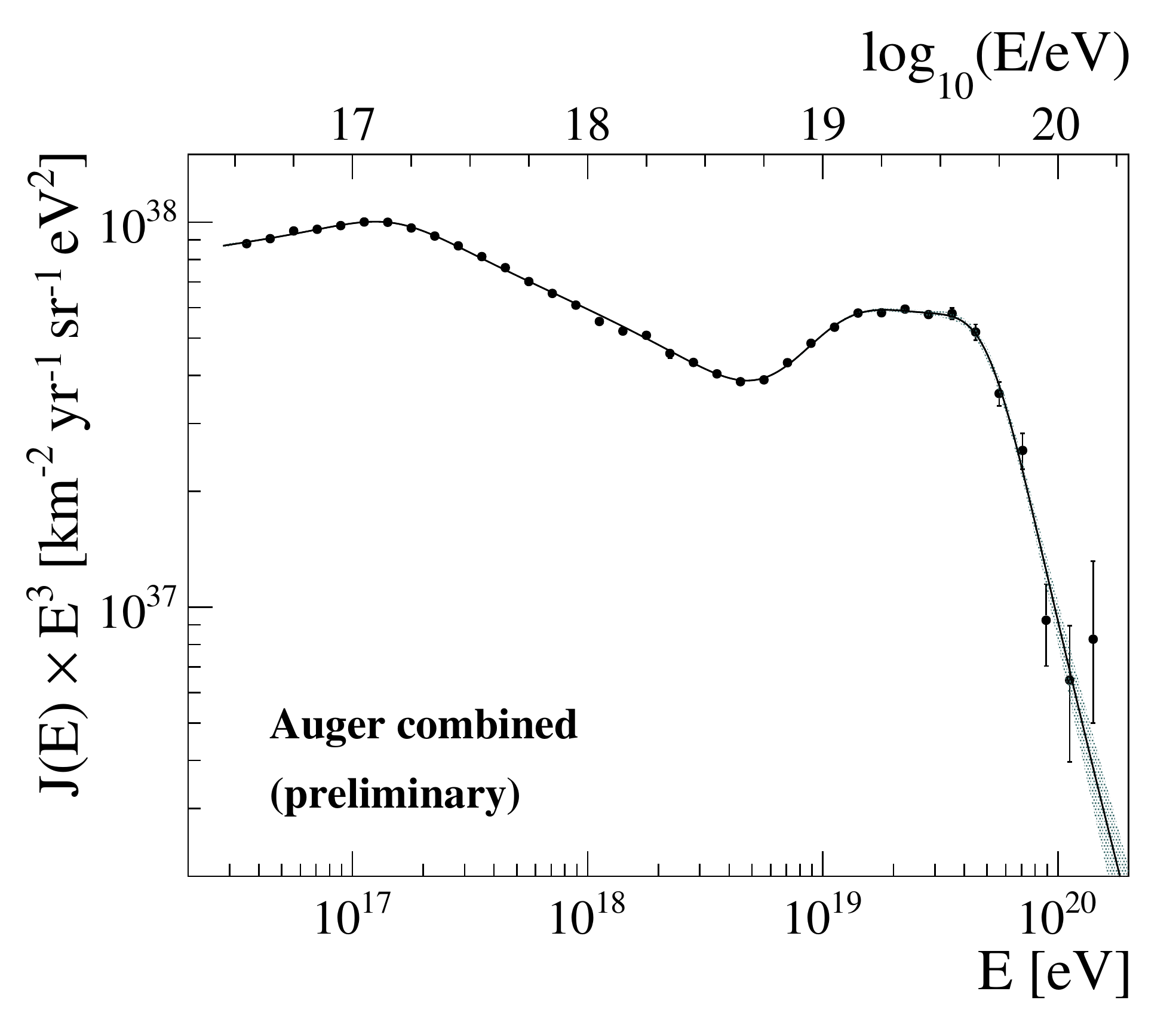}
\end{minipage}
\caption{Left: Evolution of the spectral index as a function of energy. The spectral indexes are obtained 
from power law fits to the spectrum over sliding windows of 3 bins in log(E/eV).
Right: The energy spectrum from the combination of the different  measurements.}
\label{fig:comb}
\end{figure}

At this conference, we presented for the first time the energy spectrum of cosmic rays down to 100 PeV 
using the SD750 data \cite{spec2}, obtained by exploiting a set of additional triggers allowing us to lower 
the threshold of full efficiency to $10^{17}$ eV, thus approaching the region of the so-called "2nd knee". 
Also in this case, the measurement is performed using a fully data-driven approach. The spectrum below 
the ankle cannot be correctly described by a single power law,  which is excluded at the level of $4.1\sigma$.
This conclusion is confirmed by the measurement of the showers approaching HEAT along the telescope 
axes, which, being dominated by the air-Cherenkov light emission, have a lower detection threshold \cite{found1}. 
In this analysis, the energy spectrum is derived between $10^{16.5}$ and $10^{18}$ eV with systematic 
uncertainties dominated by those in the FD energy scale and in the exposure evaluation (14\% and 15\% 
respectively). 

The spectrum obtained by combining the five data sets exploiting the total exposure of $\sim$80,000 \exprange 
is shown in the right panel of Fig.\ref{fig:comb}. 
A functional form made by a sequence of different power laws is used to fit it, as suggested by a study 
of the evolution of the spectral slope assuming a spectrum $\mathrm{dN/dE \sim E^{-\gamma}}$ and fitting 
it in sliding windows of three energy bins, as shown in the left panel of Fig.\ref{fig:comb}. 
As indicated in the plot, the spectrum is characterized by two clear 
inflection points corresponding to the second knee around $10^{17}$ eV and to the ankle near $6 \times 10^{18}$ eV.  
A new feature is visible around $10^{19}$ eV, where the spectral slope changes from $\gamma_{2}=(2.2 \pm 0.2)$ 
to $\gamma_{3}=(3.2 \pm 0.1)$.  We confirm the suppression of the flux above $\sim 5 \times 10^{19}$ eV. The
energy at which the integral spectrum drops by a factor of two below what would be the expected with no steepening
is $E_{1/2}=14 \pm 4$ EeV, at variance with that expected for uniformly distributed sources of protons, as also indicated 
by the results from our composition measurements. 

\subsection{Primary composition}

\begin{figure}
\centering
  \includegraphics[width=.9\linewidth]{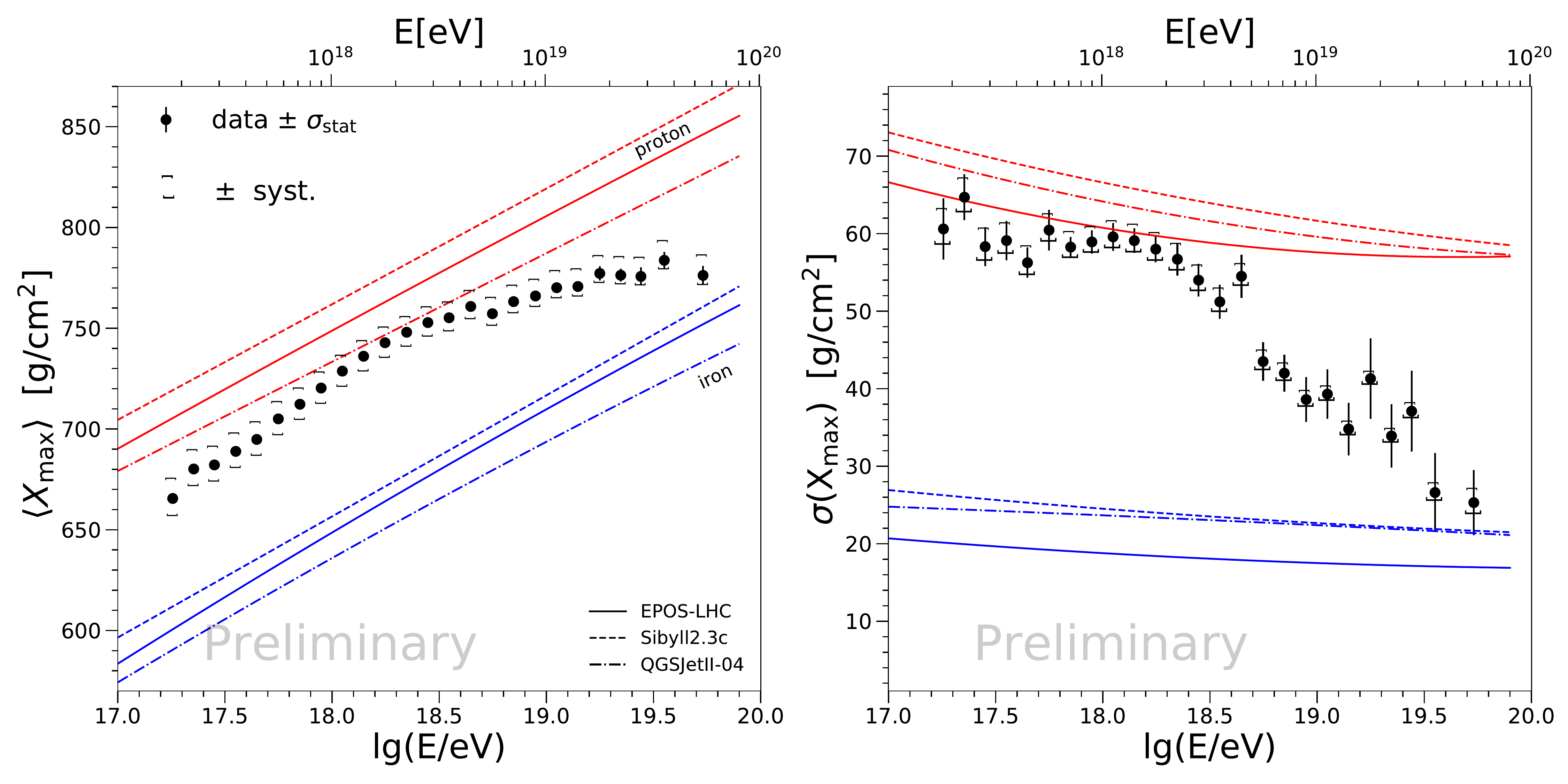}
  \caption{The evolution of  \xmaxmed (left) and \sigXmax (right) with energy in comparison with the predictions for 
  proton and iron nuclei of the post-LHC hadronic models EPOS-LHC, Sibyll 2.3c and QGSJetII-04.}
  \label{fig:mass-fd}
\end{figure}

The measurements of the depth of maximum development of the shower in the atmosphere and of its fluctuations 
together with that of the primary energy provide us with the most reliable information on the UHECRs composition.
Indeed, \xmaxmed is directly proportional to the logarithmic mass of the primary particle, while its variance
is a convolution of the intrinsic shower-to-shower fluctuations and of the dispersion of masses in the primary beam.\\
At this conference, we presented an update to this measurement based on more than 47,000 hybrid events,
more than 1000 of which above 10 EeV \cite{mass1}.  Above $10^{17.8}$ eV, this corresponds to an increase of 
20\% in statistics, while the lowest energy range is covered by the events collected by the HEAT telescopes as 
presented in \cite{mass2017}.
The evolution of \xmaxmed and its standard deviation with energy is shown in Fig.\ref{fig:mass-fd}.
%%% FIG compFD
We measure an elongation rate for \xmaxmed of $77 \pm 2 \textrm{(stat)}$  \gcmd below 
$E_{0} = 10^{18.32 \pm 0.03}$ eV and $26 \pm 2 \textrm{(stat)}$ \gcmd above this energy. 
The result is in agreement with our previous findings \cite{mass2017,masspub} and points to a composition 
getting lighter up to $E_{0}$ and going towards intermediate-heavy masses above it, at variance with the model 
expectations for constant compositions, which suggest an elongation rate of $\sim 60$ g cm$^{-2}$/decade in 
the whole energy range.
The evolution of \sigXmax  at the highest energies suggests a quite pure and heavy composition, while
in the lower energy range it is compatible with both a light or mixed one.\\
Converting the two moments of the \xmax distributions to those of the logarithmic primary mass \lnA  
as in \cite{mass2013}, we concluded that the mass is lightest at $E_{0}$, with \lnA $\sim$ 0.8 and 1.4 for 
EPOS-LHC and Sibyll2.3c  respectively. The  QGSJetII-04 model produces negative, unphysical, values 
of $\sigma^{2}$(\lnA), and as such it is unable to describe our data correctly.

Besides measuring the particle densities at different distances from the core, the SD allows us to obtain 
the distribution of the arrival times of particles at each station. The dependence of the rise-time (i.e. 
the time needed for the signal to go from 10 to 50\% of its total value) on the distance of the shower maximum 
to the ground and on the electromagnetic to muonic components of the EAS  was used in \cite{delta} to 
obtain information about the primary mass composition from SD, thus exploiting its 100\% duty cycle. 
At this conference, we presented an update of that analysis \cite{mass2} based on three more years of 
data collection and an extended  angular range of accepted events to $\theta<60^{\circ}$, obtaining 
a data sample 30 times larger than that available from the FD above 3 EeV. \\
Exploiting the correlation of the SD observable towards \xmax, we could derive the
$\left< X_\textrm {\tiny{max}}^{SD}\right>$ evolution with energy up to $E>10^{19.9}$ eV,  
although with large systematic uncertainties. The results confirm the conclusion obtained in the analysis 
of the FD data.

\begin{figure}
  \includegraphics[width=.5\linewidth]{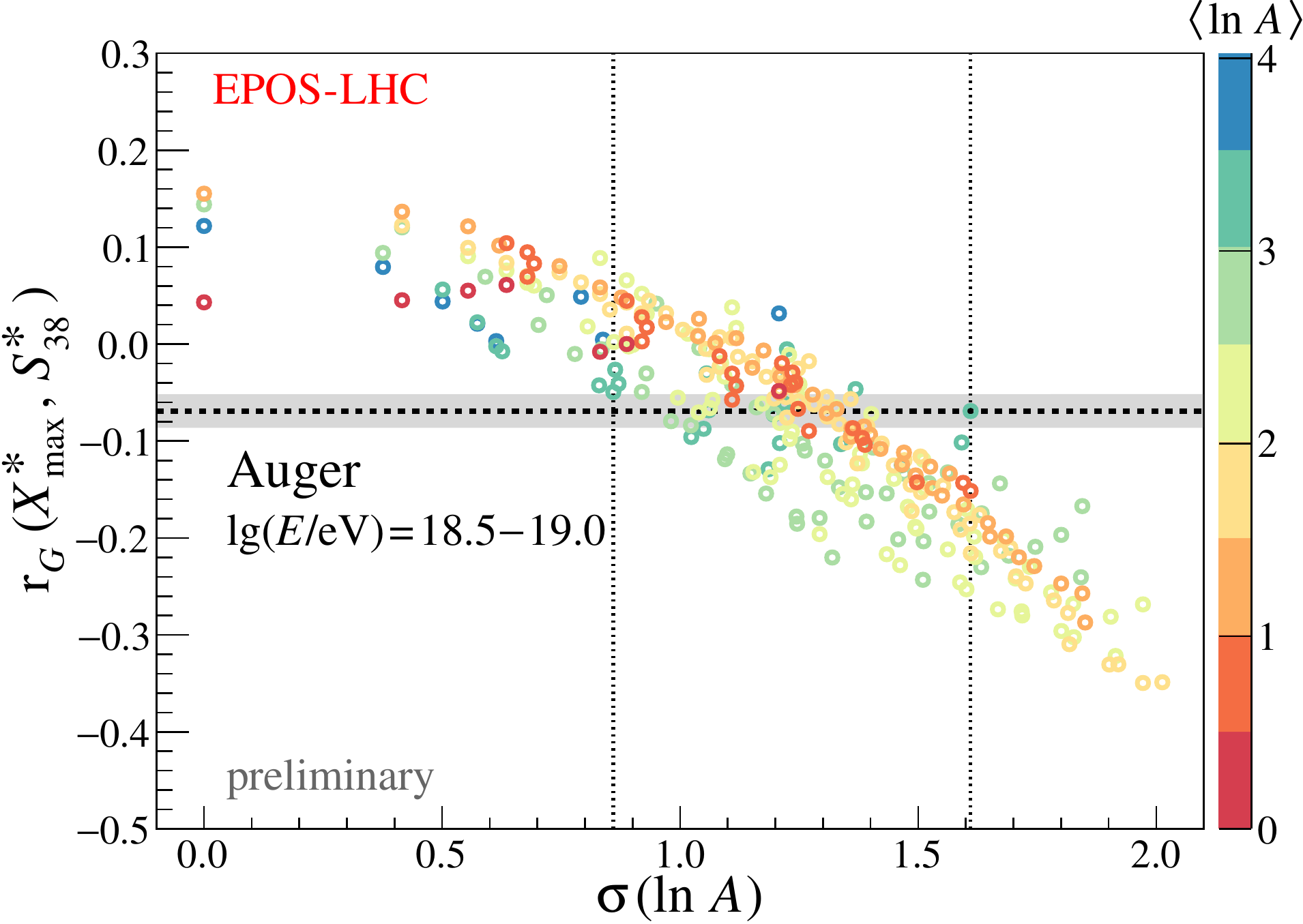}
  \includegraphics[width=.5\linewidth]{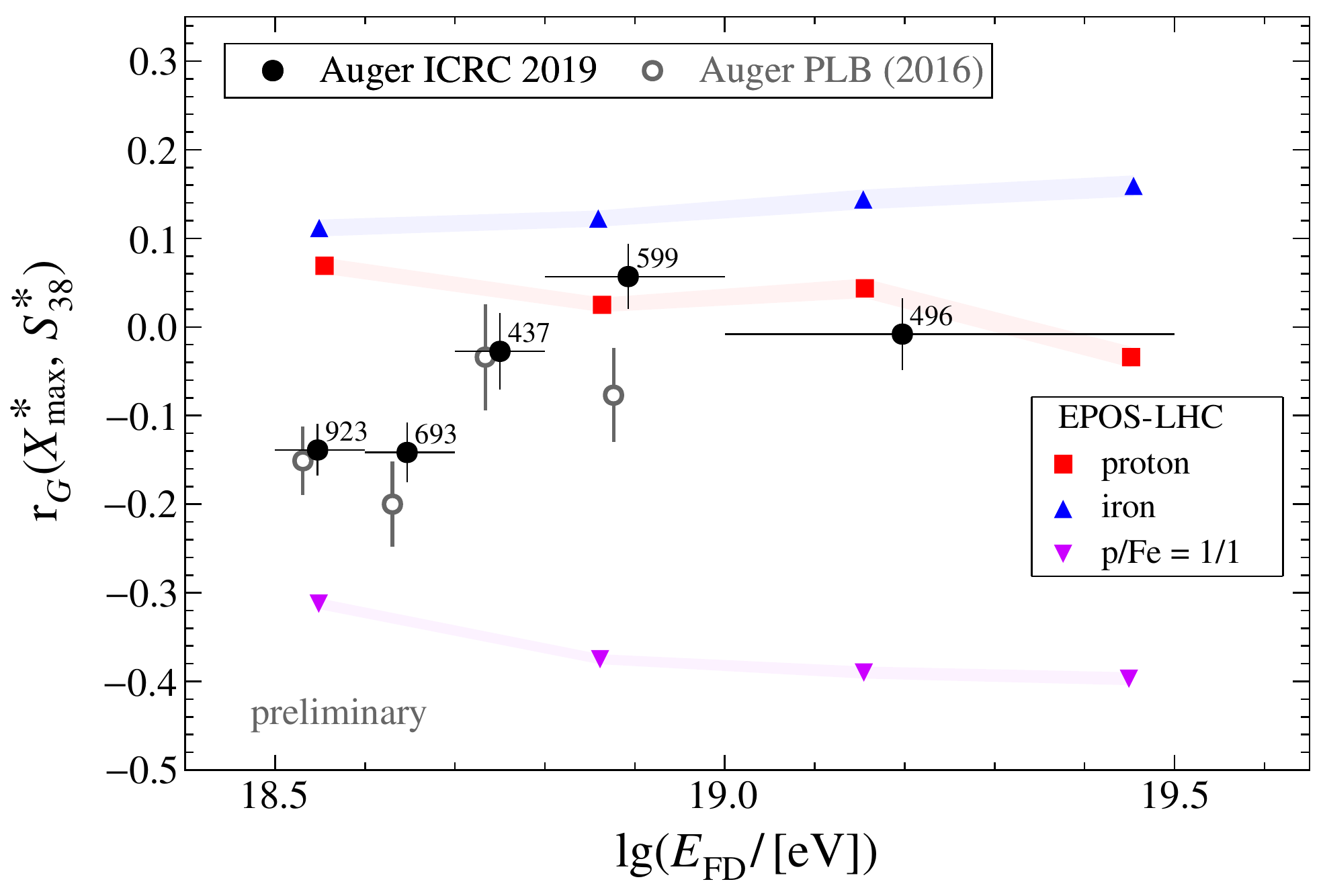}
  \caption{Left: Dependence of the correlation coefficients $r_{G}$ on $\sigma$(lnA). Right: evolution of 
  the correlation coefficient for data (full circles) in four different logarithmic energy bins. EPOS-LHC is 
  here adopted as hadronic interaction model to derive the predictions for the different compositions.}
  \label{fig:mass-cor}
\end{figure}

Hadronic showers generated by heavy primary particles develop higher in the atmosphere  and contain 
more muons with respect to those produced by lighter primaries. A study was published in \cite{massRG} 
to derive information about the mass spread in the ankle region by exploiting the correlation between 
\xmax and the signal in WCDs at 1000 meters from the core, \signal (which becomes 
more dominated by muons as the zenith angle increases).
Being based on hybrid events, this analysis is less prone to systematic uncertainties and is not affected by 
our ignorance of the hadronic interactions employed in the models. 
We presented at this conference an update of the analysis  \cite{mass1}, based on a statistics about  two 
times larger with respect to the published results, considering the energy range \erange{18.5}{19}.
To avoid a decorrelation due to the spreads of energies and zenith angles, we rescaled \xmax 
and \signal to a reference energy of 10 EeV and a zenith angle of $38^{\circ}$.  
The correlation between the two observables is expected to be positive for constant compositions in all 
models, while the experimental finding results in $r_{G}= -0.069 \pm 0.017$, with 
a systematic uncertainty $\Delta r_{G} \textrm{(sys.)}=^{+0.01}_{-0.02}$.
The dependence of the correlation coefficients $r_{G}$ on $\sigma$(lnA) is shown in Fig.\ref{fig:mass-cor}, 
where the  experimental result is indicated by the shaded grey area. 
%%% FIG Correl
The points correspond to the correlation expected in EPOS-LHC for different mixtures of (p, He, O, Fe) 
nuclei, their colour to the corresponding value of  $\left< \textrm{lnA} \right>$. The simulations are 
compatible with data in the region encompassed by the two vertical dashed lines. Similar results are 
obtained if different models are employed; in all cases, the correlation hints to a spread of masses in 
the range $0.85 < \sigma\textrm{(lnA)} < 1.6$. 
As shown in the right panel of Fig.\ref{fig:mass-cor},  the correlation at energies \erange{18.5}{18.7},
below the ankle, is significantly different from zero (6.4$\sigma$), while smaller mixings are allowed 
above this range. In the considered energy range \erange{18.5}{19.0}, pure compositions and compositions 
consisting of only protons and helium can be excluded at $5 \sigma$ level.  

\begin{figure}
\begin{subfigure}{.6\textwidth}
  \centering
  \includegraphics[width=0.9\linewidth]{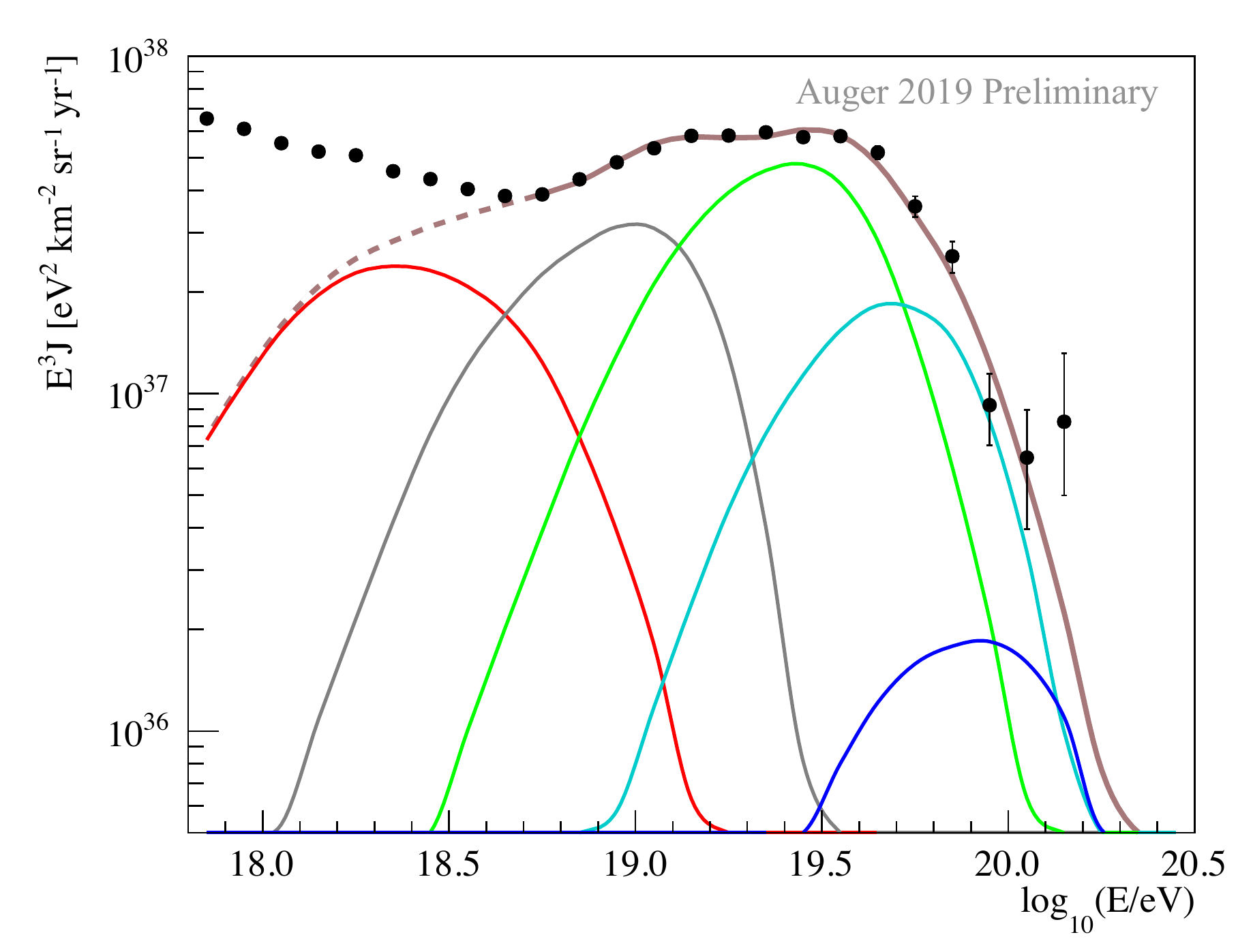}  
\end{subfigure}
\begin{subfigure}{.4\textwidth}
  \centering
  \includegraphics[width=0.85\linewidth]{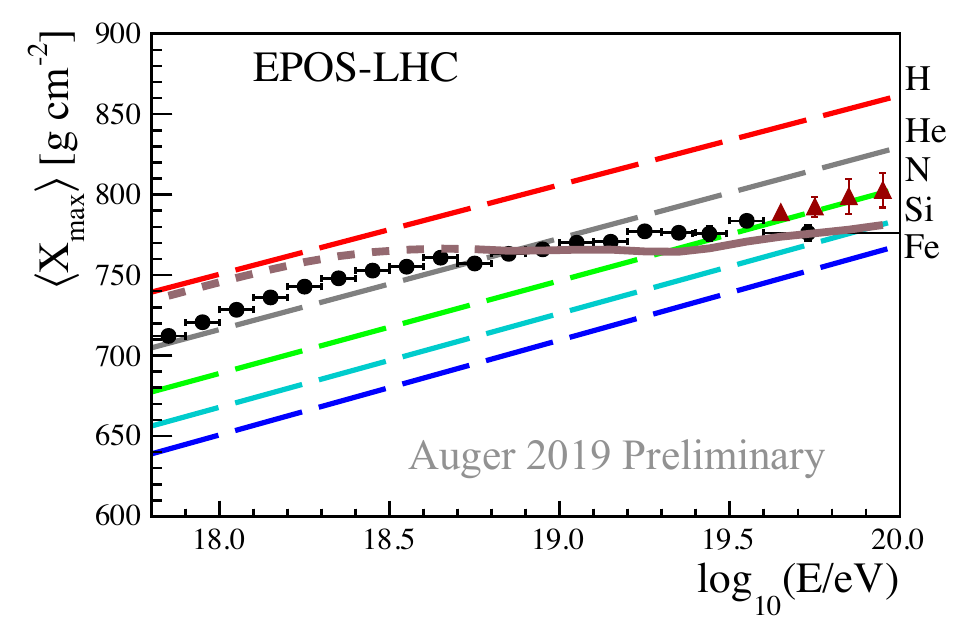}  
  \includegraphics[width=0.85\linewidth]{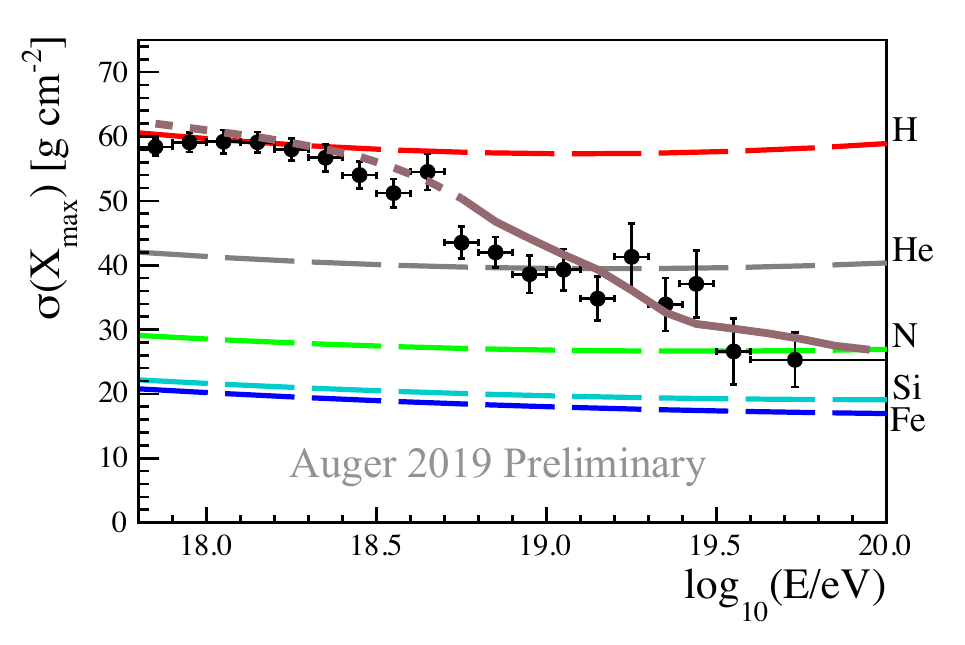}  
\end{subfigure}
 \caption{Left: The simulated energy spectrum (multiplied by $E^{3}$) at the top of the Earth's atmosphere 
 obtained with the best fit parameters (see text): all-particle (brown curve), $A=1$ (red), 
  $2 \leq A \leq 4$ (grey), $5 \leq A \leq 22$ (green), $23 \leq A \leq 38$ (cyan), $A \geq 39$ (blue). The 
  combined energy spectrum as measured by Auger (Fig.\ref{fig:comb}, right) is shown for comparison
  with the black dots.
  Right: The first two moments of the \xmax distributions as predicted for the model (brown curve) versus
  pure compositions. Only the energy range indicated by the solid brown line is included in the fit. The
  measured mean \xmaxsd are shown with purple triangles for comparison.
}
\label{fig:cf}
\end{figure}

An astrophysical interpretation of the Auger data on the energy spectrum and mass composition can be 
obtained starting from different assumptions on the possible sources. 
By assuming the same model of \cite{combfit},  in which the sources of UHECRs are of 
extragalactic origin and accelerate nuclei in electromagnetic processes with a rigidity-dependent maximum 
energy, $E_\textrm{max}(Z) = E_\textrm{max}(p)/Z$,
a preliminary comparison of the energy spectrum and composition expected 
at the top of the Earth's atmosphere  to the measured ones can be obtained in a combined fit procedure. 
As shown in Fig.\ref{fig:cf}, even using this simple model we reached the statistical power 
and the knowledge of systematic uncertainties such to be able to reproduce the different features of 
the spectrum and the average behaviour of the \xmax and its  standard deviation.  
The best-fit parameters hint to a quite hard injection spectrum and a low rigidity cut-off for protons, thus 
implying a non-negligible fraction of heavier elements at the highest energies.  
Within the systematic uncertainties, the results agree with those previously published \cite{combfit}. 
%%% FIG CombFit

\subsection{Multi-messengers} 

\begin{figure}
  \includegraphics[width=.5\linewidth]{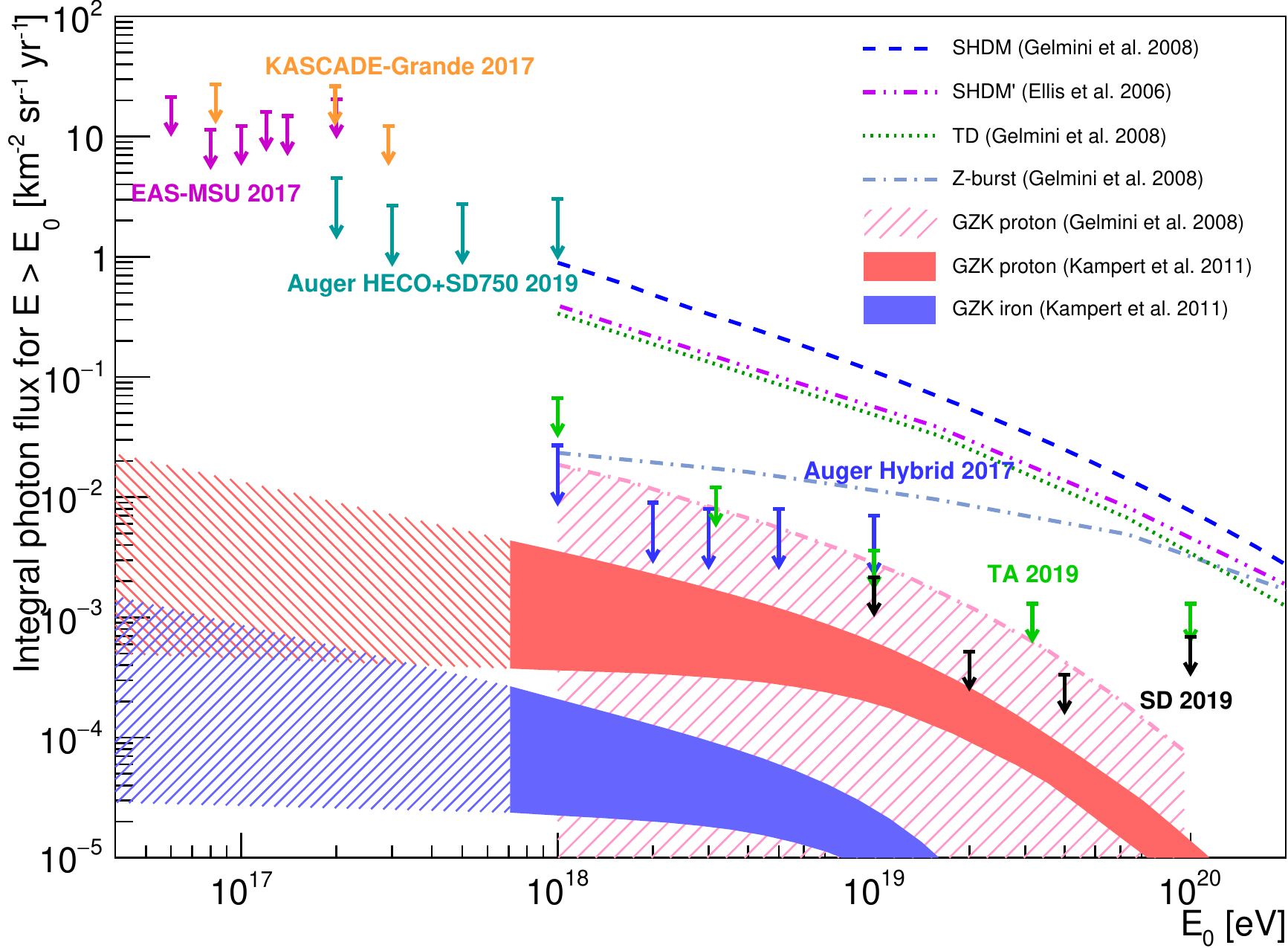}
  \includegraphics[width=.5\linewidth]{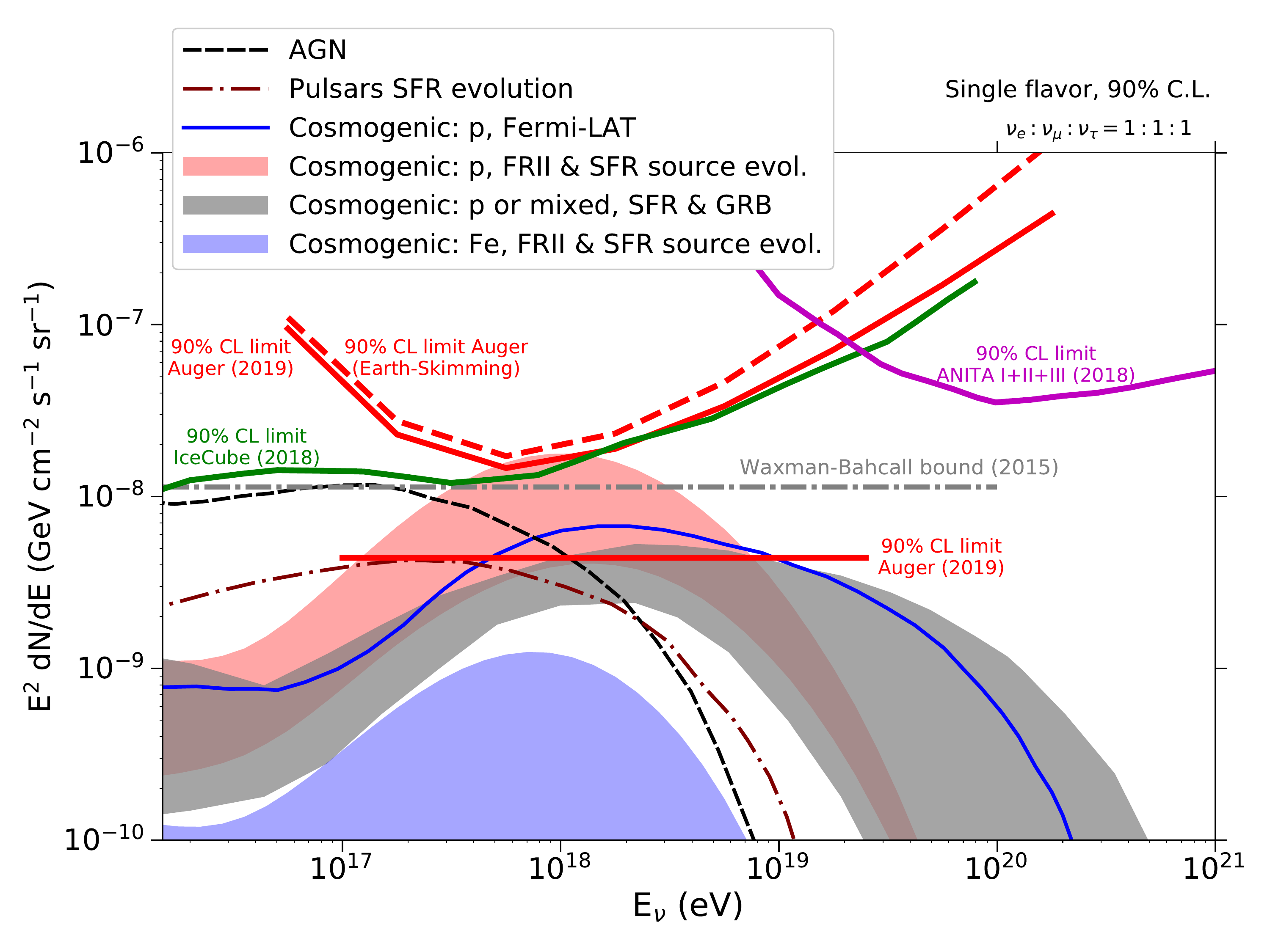}
\caption{Upper limits on the cosmogenic photon (left) and neutrino (right) fluxes compared to limits 
from other experiments and to model predictions (\cite{neut2,neut1} and refs.therein).}
  \label{fig:diffuse}
\end{figure}

With its high potential in the identification of photons and neutrinos, the Pierre Auger Observatory has a key 
role in the field of multi-messenger astrophysics. Cosmogenic neutrinos and photons are produced in the 
interactions of UHECRs in the  background radiation fields during their propagation from their sources 
to Earth. 
Neither are deflected in the Galactic or intergalactic magnetic fields; they point back to their sources, 
making them ideal messengers in targeted searches. While photons cover travel distances of the order of 
$\sim 4.5$ Mpc at 1 EeV, neutrinos allow us to probe sources up to cosmological distances.
Their fluxes depend on the properties of the sources and on the composition of the primary beam. A
copious production of $\nu$ and $\gamma$ can be expected, e.g., if the suppression of the UHECR flux above 
a few tens of EeV is due to propagation effects and if the proton component is dominant. On the contrary, 
an explanation of the cutoff as due to the exhaustion of the sources would lead to much lower fluxes of
neutral particles and would point to a mixed composition.\\
The selection of photons in Auger is based on the fact that photon-induced showers present a more 
elongated profile in the atmosphere, and thus a larger \xmax, a steeper lateral distribution, causing the 
involvement of a lower number of SD stations in the events, and a reduced production of muons with respect 
to hadronic showers \cite{neut2}.\\
Neutrinos are looked for based on the selection of horizontal showers. 
In hadronic-induced showers above $\sim 60^{\circ}$, the electromagnetic component is indeed almost 
completely absorbed in the atmosphere, and only muons are detected in the SD. On the contrary, in the case 
of neutrino events with similar arrival directions, the first interaction would happen lower 
in the atmosphere, producing a considerable amount of electrons and photons at the ground. Two main 
categories of events are considered: Earth-skimming, induced by $\nu_{\tau}$ travelling from below the 
Earth crust in directions between $90^{\circ}$ and $95^{\circ}$ and producing a $\tau$ lepton which can 
then generate a shower above the SD, and downward-going events due to neutrinos of any flavour. 
Thanks to the larger grammage and to the density of the Earth with respect to the atmosphere, the first 
is the most sensitive channel \cite{neut1}.\\
The current upper limits on the cosmogenic fluxes of photons and  neutrinos are shown in Fig.\ref{fig:diffuse}.

The energy range covered in the search for diffuse photons is here extended to about three decades, 
thanks to an exposure now reaching $40,000$ \exprange, and to the analysis of data from the low energy 
enhancements of the Observatory, namely SD750 and HEAT. The limits above EeV energies put strong 
constraints on top-down models of UHECRs origin \cite{topdown} and on the most optimistic models of photon 
production from the interaction of protons with the CMB.
%%% FIG mm

\begin{figure}
  \includegraphics[width=.53\linewidth]{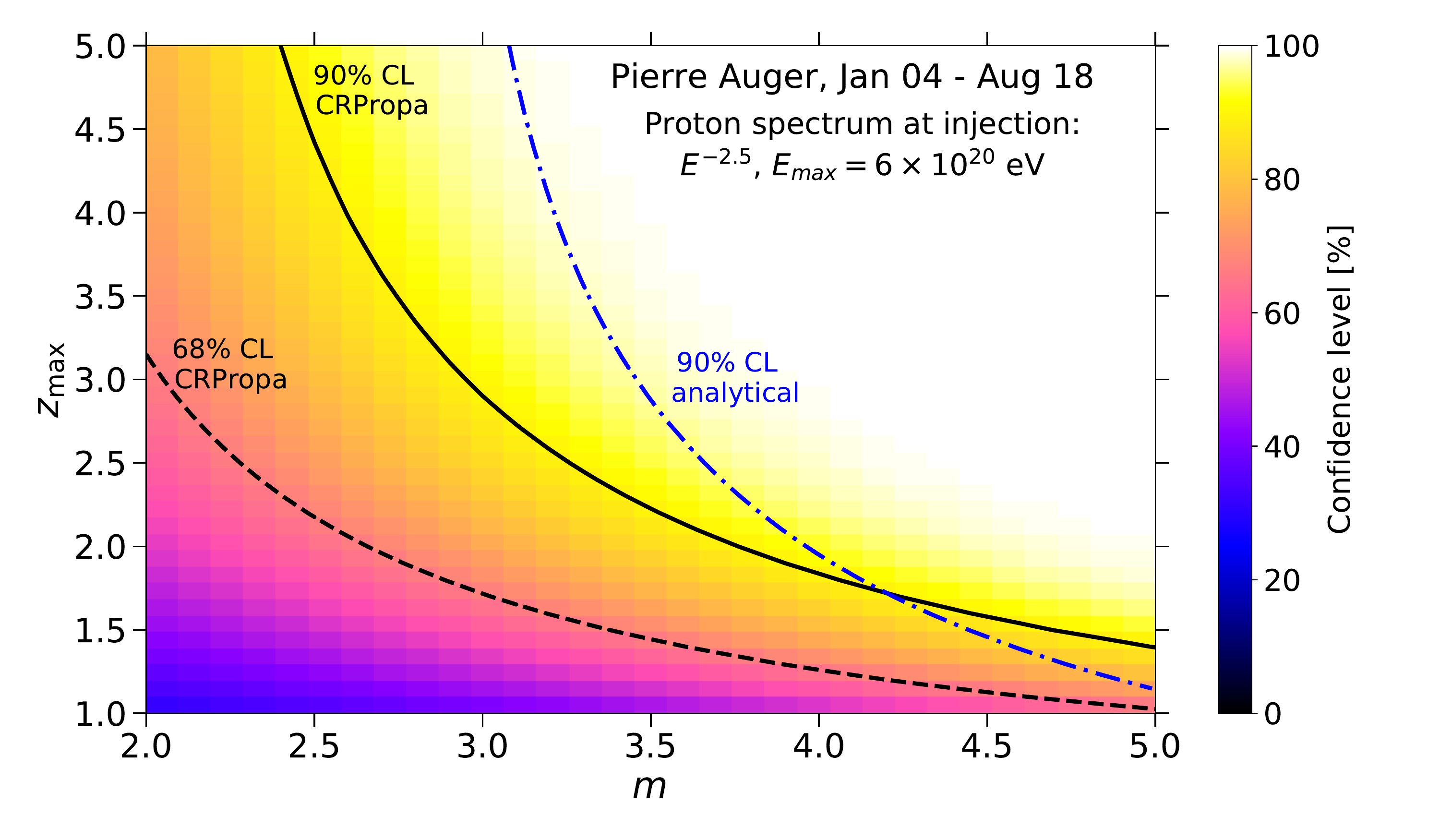}
  \includegraphics[width=.47\linewidth]{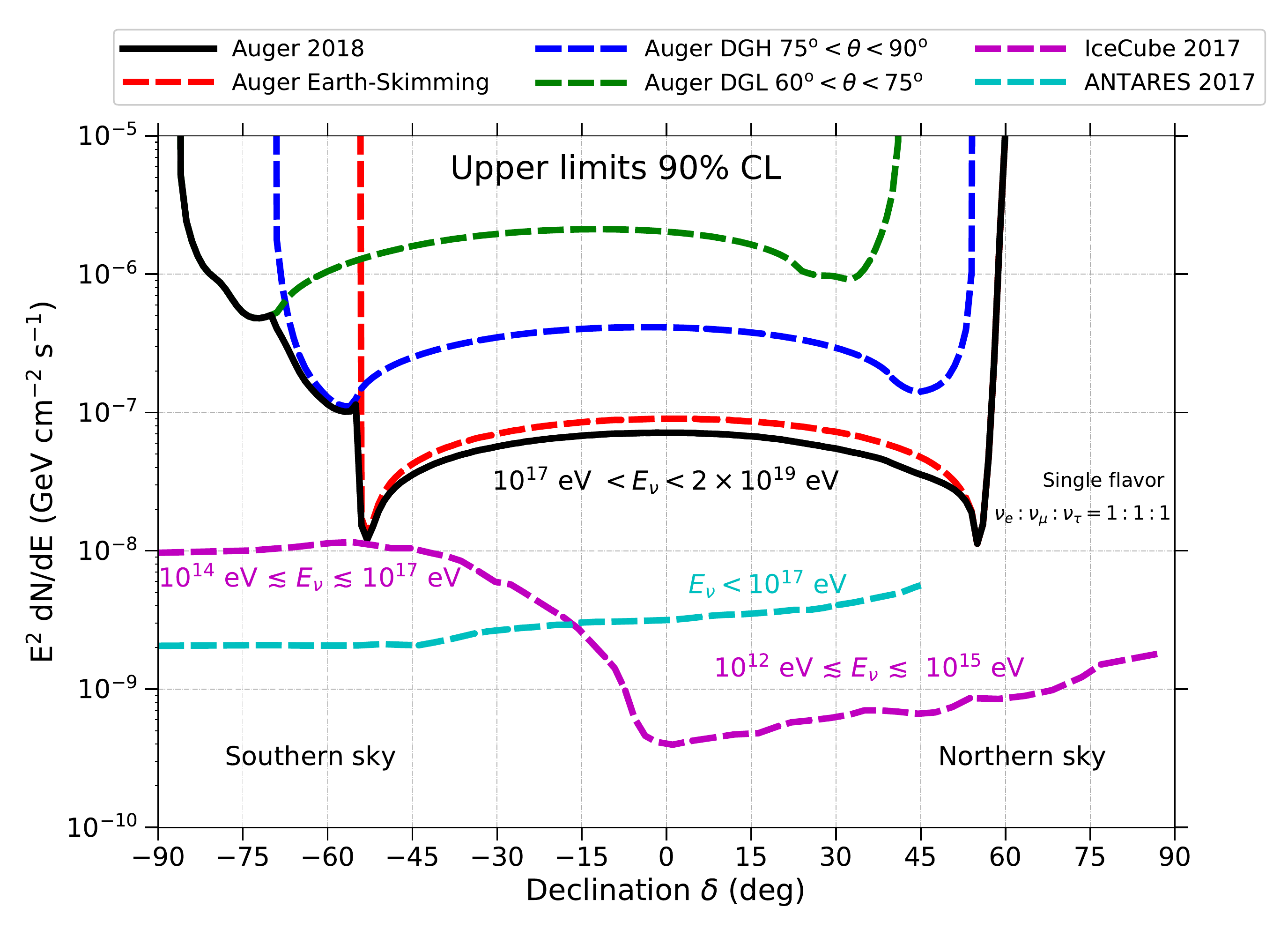}
\caption{Left: Constraints on UHECR source evolution models (see text). Right: Upper limits at 90\% C.L. on 
the normalization $k_{PS}$ of a single flavour point-like flux of UHE neutrinos
$dN/dE = k_{PS}E^{-2}$ as a function of the source declination $\delta$.}
  \label{fig:nusteady}
\end{figure}

The single flavour 90\% upper limit on the integrated neutrino flux, for an assumed  flux $\sim E_{\nu}^{-2}$ 
and no neutrino candidates, is $k_{90} < 4.4 \times 10^{-9}$ \frange in the energy range 
$10^{17} - 2.5 \times 10^{19}$ eV. 
The differential upper limit characterizes the energy dependence of the sensitivity of a neutrino experiment: 
as shown in the right panel of Fig.\ref{fig:diffuse}, with the Auger Observatory the best sensitivity is achieved 
for energies around 1 EeV. Describing  the evolution of the sources by  $\Psi(z) \propto (1+z)^m$, up to a maximum 
redshift $z_\textrm{max}$, we could perform a scan in the $(z_\textrm{max},m)$ phase space and obtain the expected 
flux and the number of cosmogenic neutrino events in Auger (using the propagation code CRPropa  
\cite{propa}) for each configuration. 
From the non-observation of neutrinos, we could thus exclude a significant part of the parameter space, 
as shown in the left panel of Fig.\ref{fig:nusteady}.

Besides diffuse neutrino fluxes, exploiting the very good angular resolution of the Observatory (less than 
$0.5^{\circ}$ for $\theta > 60^{\circ}$) we can look for steady sources of UHE neutrinos. Upper bounds on 
the normalization of a differential neutrino flux $E_{\nu}^{-2}$ as a function
%%% FIG steady
of declination are shown in the right panel of Fig.\ref{fig:nusteady}. The energy range of Auger is different 
and complementary to those of IceCube and Antares \cite{ic,ant}.

At this conference, we also presented  the results of our studies on transient sources of UHE neutrinos of 
the order of an hour or less. The sensitivity of the SD to these kind of sources depends not only on the 
declination, as in the case of steady sources, but also on the efficiency of the detection during the time 
occurrence of the transit. For sources in transit within the field of view corresponding to the zenith angle 
range  $\theta=(60^{\circ} - 95^{\circ})$, where the SD is most sensitive to UHE neutrinos, the sensitivity 
is unrivaled with respect to other neutrino detectors.\\
At the time of the GW170817, detected by LIGO and Virgo \cite{gw}, the source was located in the most 
sensitive part of the Auger field of view, and we provided the strongest constraints on the UHE neutrino 
fluence from the merger \cite{paogw}.
Two other types of potential transient sources were scrutinized in \cite{mm1}. Binary black hole mergers 
(BBH), detected by the gravitational wave detectors, could in principle be sources of neutrinos, if  
matter debris and strong magnetic fields are present in the source environment \cite{bbh}. 
Combining searches for neutrinos from 21 BBH mergers, and considering an emission spectrum 
$E^{-2}$, the non-observation of any candidates in the 24 hours after the mergers brought to an upper
limit on the luminosity of a universal isotropic source of UHE neutrinos of $L_\textrm{up} <  4.3 \times 10^{46}$ 
erg/s, with maximum sensitivity  about 14.6 hours after the merger.\\
UHE neutrino events from the blazar  TXS 0506+056, for which a correlated emission of neutrinos and 
photons was found \cite {icTXS,icTXS2} below the energy threshold of Auger, were also searched for 
in two different time windows correlated to the neutrino excess found with IceCube and to the following 
associated high-energy neutrino alert, but no UHE neutrinos have been observed. Despite the smaller 
sensitivity to neutrinos from the TXS 0506+056 flares with respect to IceCube, due to the periodic visibility 
with interruptions, we underline that we are scrutinizing a different higher energy range, unreachable by 
other observatories.

\subsection {Arrival directions} 

In \cite{dipole}, the Pierre Auger Collaboration announced the observation of a large-scale anisotropy in 
the arrival directions  of cosmic rays above $8 \times 10^{18}$ eV.  At this conference, this analysis was 
extended to cover more than three decades in energy, exploiting 
about 15 years of data \cite{arr2}. For energies above 4 EeV, data from the fully efficient SD1500 array 
were used, reaching a total exposure of $92,500$  and $60,700$ \exprange for events with 
$\theta \leq 80^\circ$ and $\theta \leq 60^\circ$ respectively. 

\begin{figure}
\includegraphics[width=.60\textwidth]{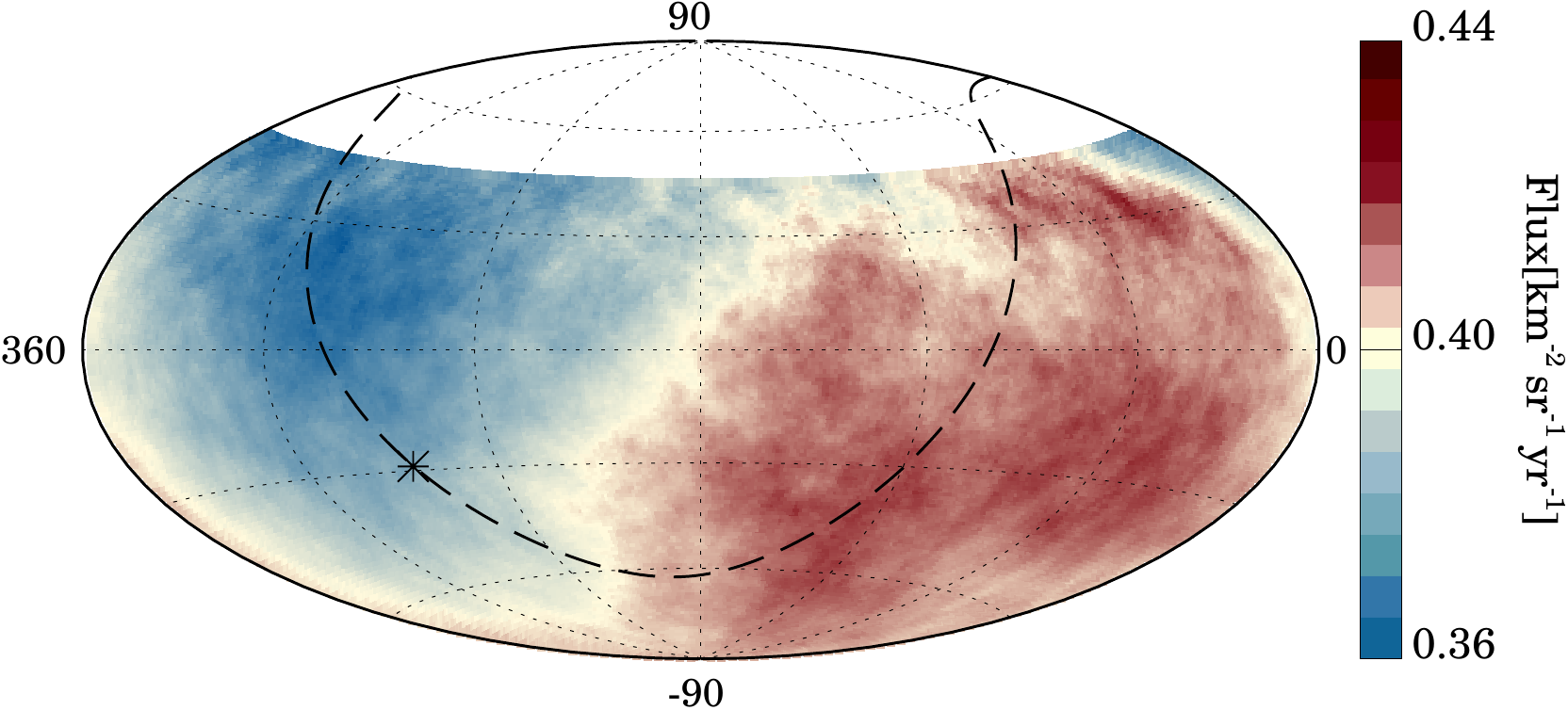}
\includegraphics[width=.40\linewidth]{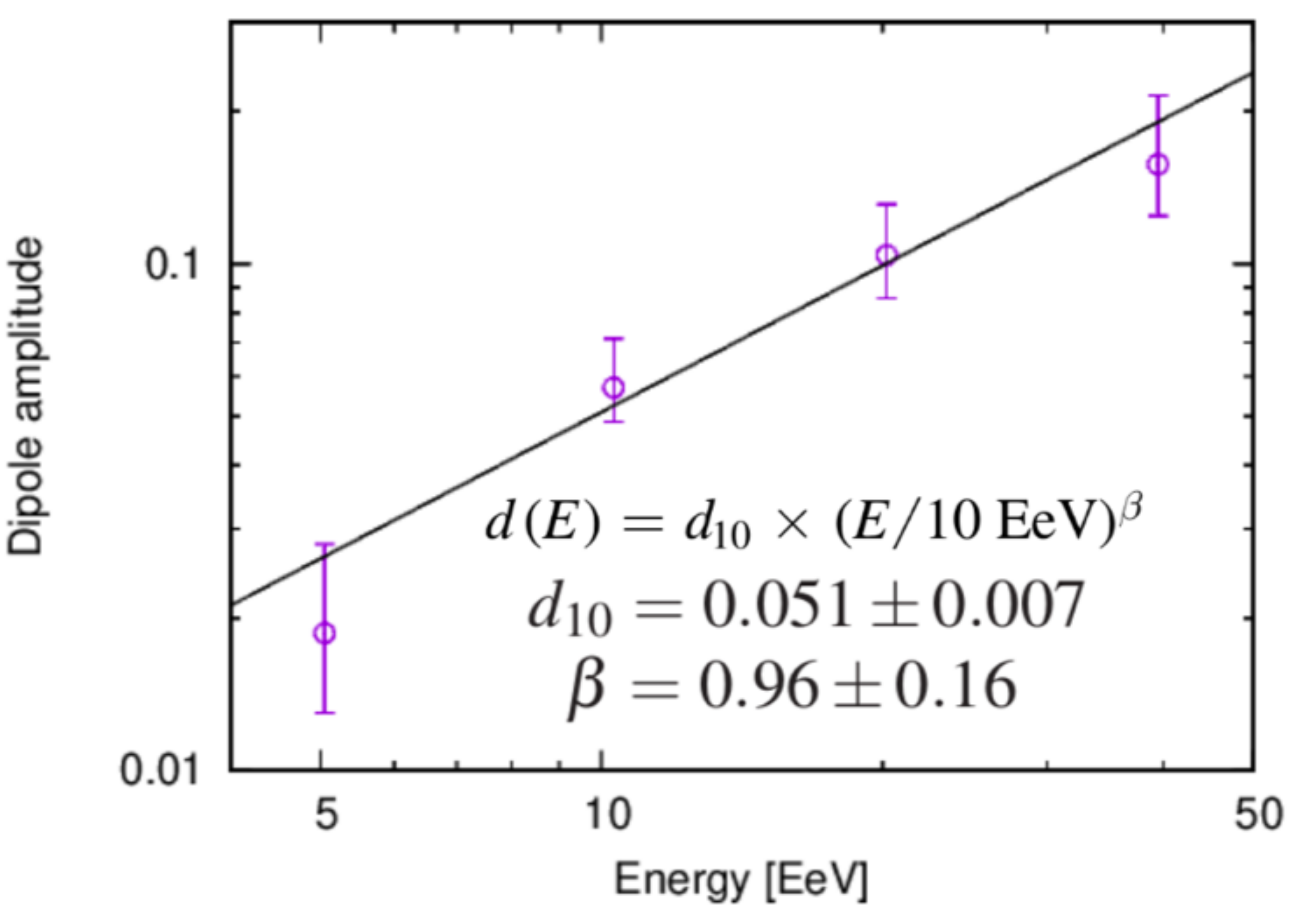}
\caption{Left: The CR flux above 8 EeV, averaged on top-hat windows of $45^{\circ}$ radius 
(equatorial coordinates).
The Galactic plane and the Galactic center are indicated by a dashed line and a star respectively. 
Right: Energy dependence of the dipolar amplitude measured in four energy bins above 4 EeV.}
\label{fig:LSA}
\end{figure}

Under the assumption that higher multipoles are negligible, we found a total dipolar amplitude for 
$E \geq 8$ EeV  of $d=0.066 \pm 0.012$, pointing $\sim 125^{\circ}$ away from the direction of the 
Galactic center, as such indicating an extragalactic origin of the modulation. The overall flux in this 
energy bin is shown in the left panel of Fig.\ref{fig:LSA}.
%%% FIG Dipolo1 e dipolo2
The possible presence of quadrupolar components has been considered, but their amplitudes 
result to be negligible, and even in this case the dipole is consistent with that found by neglecting 
higher-order multipoles.\\
The same study, free from assumptions on the shape of the underlying angular distribution, has been 
performed considering a full-sky search for large-scale anisotropies, in a joint working group of the Pierre 
Auger and Telescope Array collaborations \cite{joint5}. By a careful reweighting of the exposures and  
cross-calibrating the respective energy scales, the data from the two experiments could be analyzed together. 
The dipolar amplitude found is again consistent with that obtained by Auger alone,
with smaller uncertainties when allowing for non-vanishing quadrupole moments.

Thanks to the large exposure, with the Auger Observatory we studied the evolution of the dipolar amplitude with energy, 
which is shown in the right  panel of Fig.\ref{fig:LSA}. The linear growth can be reproduced with 
$d = d_{10}(E/10 \ \textrm{EeV})^{\beta}$, where $d_{10} = 0.051 \pm 0.007$
and $\beta = 0.96 \pm 0.16$; an evolution of the amplitude constant with energy is disfavoured at the 
level of $5.1 \sigma$ by a likelihood ratio test.  A study of the effect on the extragalactic dipole of the 
deflections caused by the Galactic magnetic field showed that in this case, depending on the particle 
rigidities, the dipole direction would tend to align in the direction of the spiral arms and suffer a 
reduction of the amplitude \cite{dipole2}.

\begin{figure}[t]
  \includegraphics[width=.5\linewidth]{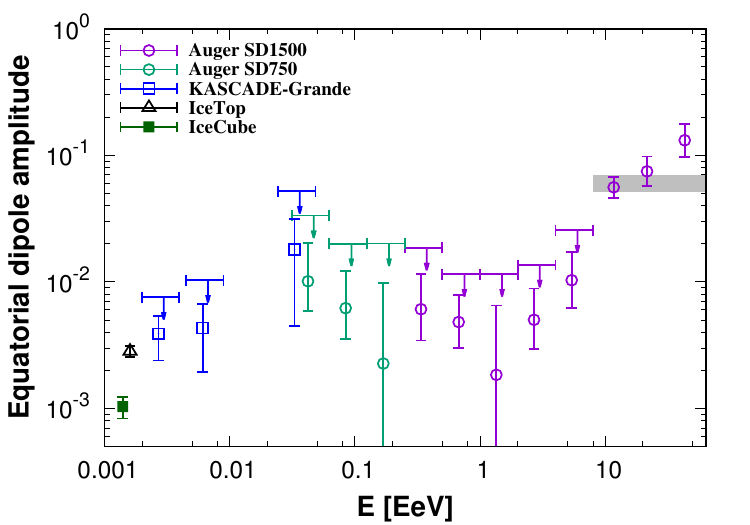}
  \includegraphics[width=.5\linewidth]{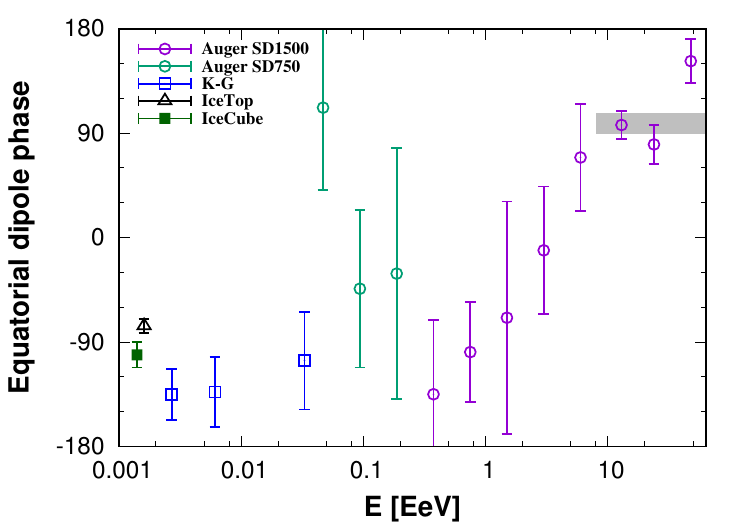}
\caption{Amplitude (left) and phase (right) of the equatorial dipole amplitude in the extended 
energy range. The upper limits $d_{\perp}^{UL}$ are indicated by arrows for the bins where 
$d_{\perp}<d_{\perp}^{99\%}$ (see text).
The results at lower energy from IceCube, IceTop and KASCADE-Grande (\cite{arr2} and 
refs. therein) are also shown.}
\label{fig:leRange}
\end{figure}

Further studies have been conducted to extend the search for large-scale anisotropies to a lower energy 
range, exploiting the data from the infill array SD750 and using the East-West method \cite{ew}, less 
sensitive to systematic effects that could result in spurious anisotropies, in place of the standard Rayleigh 
analysis. The amplitudes and phases of the equatorial dipole in the extended energy range are 
shown in Fig.\ref{fig:leRange}.
There, we also indicate with the 99\% CL upper bounds the result in the bins where the measured amplitude 
is smaller than the value within which 99\% of the simulations with isotropic distributions of the same 
number of events would fall. Although the amplitudes at low energy are not significant, the phase of the 
equatorial dipole appears to change from values pointing close to the right ascension of the Galactic 
center below EeV energies towards the opposite direction, again indicating an extragalactic origin for the 
dipolar anisotropy above a few EeV.

At higher energies, where the rigidities of the particles are high enough and cosmic rays from far sources 
are suppressed due to the energy losses they suffered during propagation, searches for 
intermediate-scale anisotropies have been performed by exploiting 15 years of data, reaching an 
unprecedented exposure of $\sim 100,000$ \exprange in a declination range between $-90^{\circ}$ and 
$+45^{\circ}$ \cite{arr3}. 
%%% FIG leLSA
A model-independent blind search for overdensities was performed over the whole field of view, in a wide 
energy threshold range $E_\mathrm{thr}=32-80$ EeV using a 1 EeV step and scanning circular regions 
$\Psi$ from $1^{\circ}$ to $30^{\circ}$ in steps of $1^{\circ}$.
The most significant excess was found for $E > 38$ EeV in an angular window of $27^{\circ}$ centered 
in the direction $(\alpha=202^{\circ}, \delta=-45^{\circ})$, a region densely populated by extragalactic 
objects among which there is the closest radio-Galaxy Centaurus A. Indeed, the excess of events in 
its direction, already pointed out in the past \cite{cenA}, reached a one-sided post-trial significance 
of $3.9 \sigma$ for energies above $E_\mathrm{thr}=37$ EeV and $\Psi=38^{\circ}$.

The arrival directions of UHECRs were also studied against the distribution of nearby extragalactic 
matter traced by different catalogs of possible sources up to a distance of $\sim 250$ Mpc. 
Similarly to \cite{catalogs}, we considered (a) a sample of 32 starburst Galaxies (SBGs)
selected based on their continuum emission at 1.4 GHz, used as a proxy for their UHECR flux; 
(b) a selection of $\gamma$AGNs from the 3FHL catalog, weighting the sources with the integral flux 
from 10 GeV to 1 TeV; (c) the 2MRS catalog of nearby matter farther than 1 Mpc; (d) a selection of 
Swift-BAT radio-loud and radio-quiet AGNs.\\
%%% FIG correl
\begin{figure}
  \includegraphics[width=.52\linewidth]{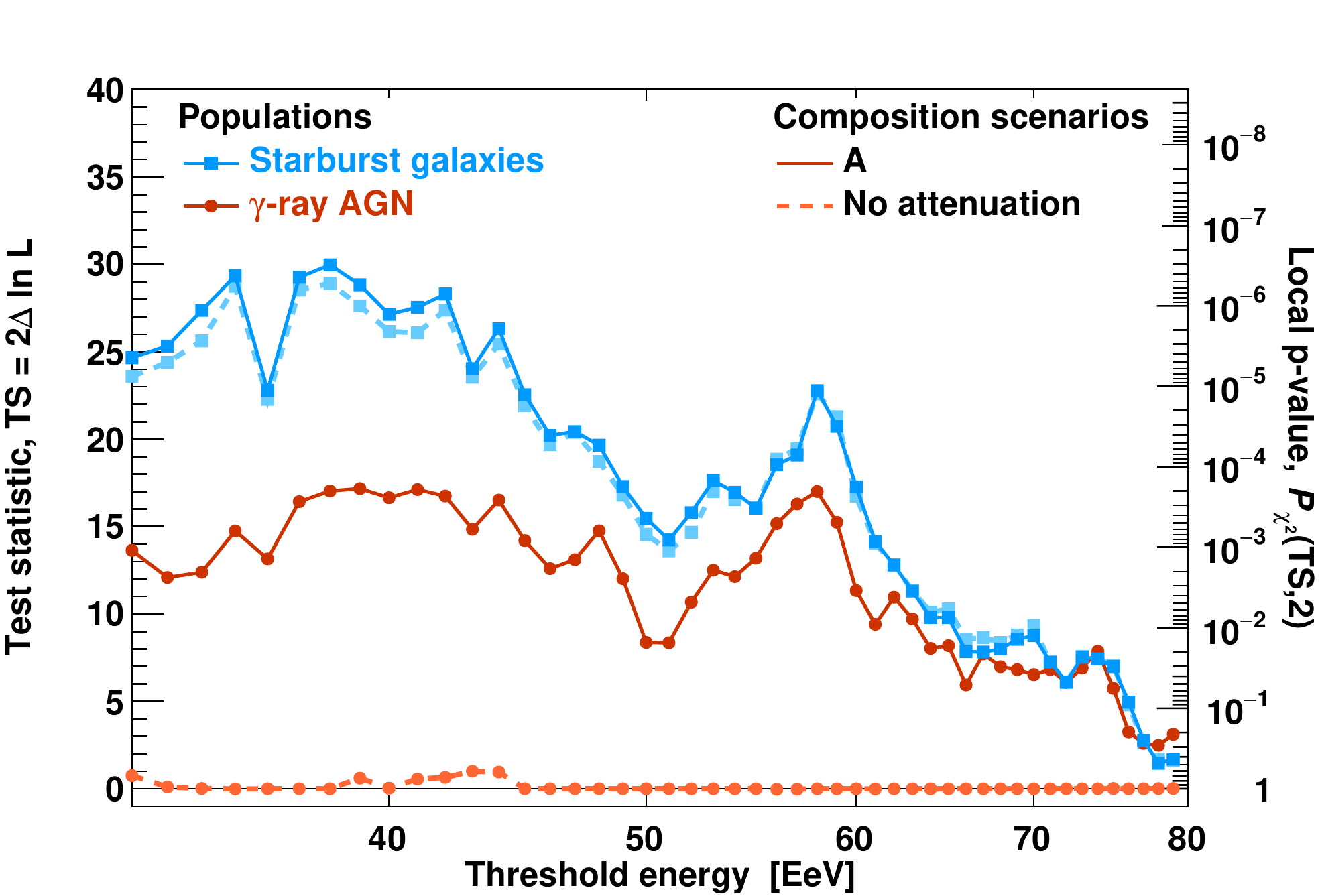}
  \includegraphics[width=.48\linewidth]{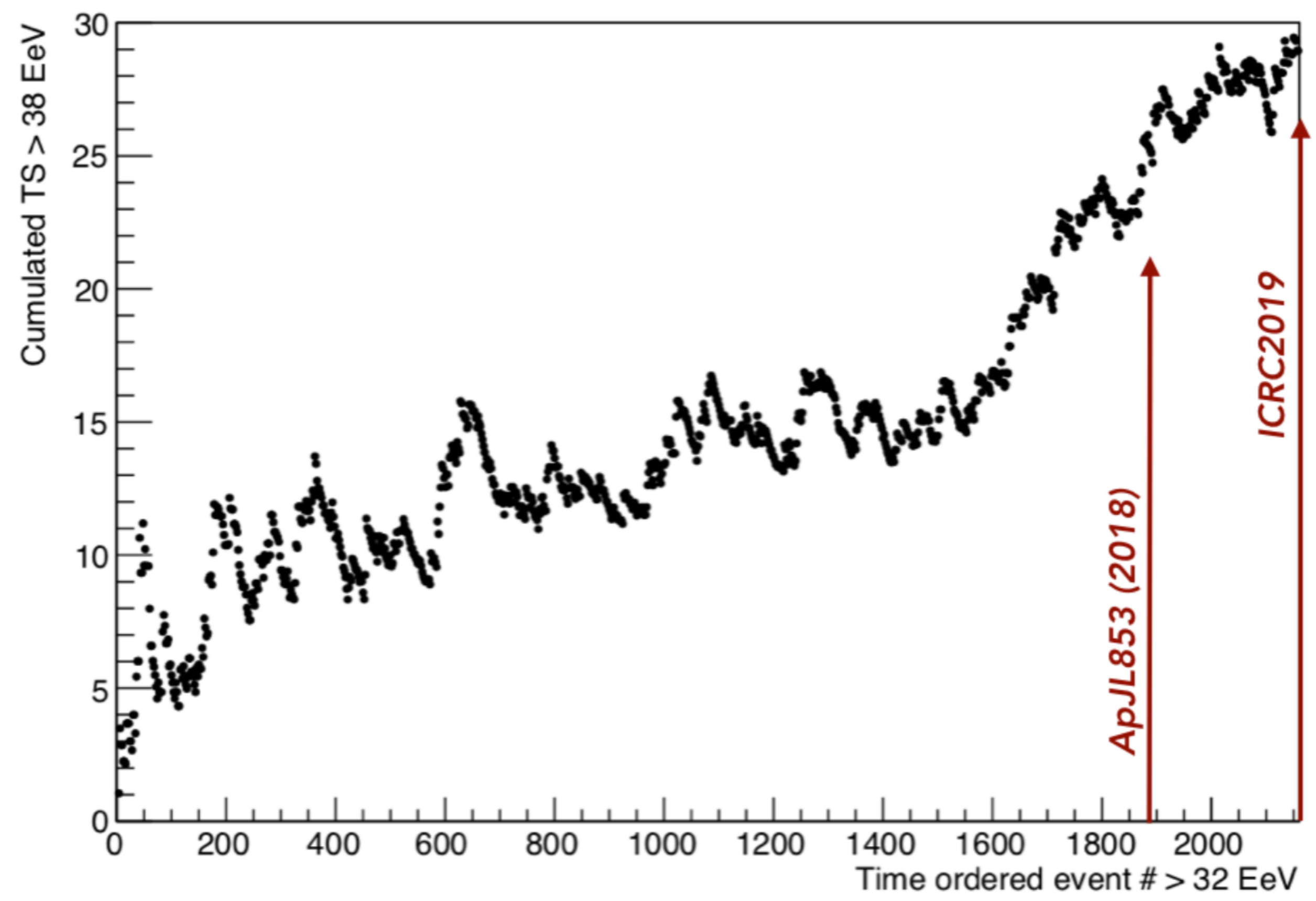}
\caption{Left: Maximum likelihood-ratio as a function of energy threshold for the models based on SBGs 
and $\gamma$AGNs. The results are shown in the attenuation (full line) and no-attenuation (dashed line) 
scenarios. Right: Cumulated test statistics for $E_\mathrm{thr}=38$ EeV as a function of the time ordered 
number of events (for the SBG-only model). 
The number of events at the time of \cite{catalogs} and of this conference are indicated 
by the red arrows.}
\label{fig:catcorr}
\end{figure}
In this analysis, we exploited not only the source directions but also the possible ansatzes for the 
relative UHECR fluxes from source candidates provided by the observations from Fermi-LAT. 
Furthermore, our current knowledge of the cosmic ray composition could be used to evaluate the 
attenuation in intensity due to the energy losses during propagation. \\
 A likelihood ratio analysis allowed us to evaluate the smearing angle and the fraction of anisotropic 
 cosmic rays for each case.
The test statistics obtained for the correlation with SBGs and $\gamma$AGNs is shown in the left panel
of Fig.\ref{fig:catcorr}. 
 The highest significance is obtained for the case of SBGs, where we found an anisotropic fraction 
 of $(11 \pm 5) \%$ of events with energy above 38 EeV in a smearing angle of $15^{\circ}$.  Note that
 the attenuation of particles in the case of the nearby SBGs is much less important that for more distant blazars. 
After penalization for the scanning in energy,  the significance of the correlation results to be $4.5 \sigma$.
The cumulated test statistics for $E_\mathrm{thr}=38$ EeV  is shown in the right panel 
of Fig.\ref{fig:catcorr} as a function of the number of collected events.
\newpage

\section{Hadronic interactions}
\label{sec:hep}
%intro 
The interpretation of the experimental observables in terms of primary composition is prone to systematic 
uncertainties, mainly due to the lack of knowledge on hadronic interactions at ultra-high energies.  
On the one hand, additional data from collider and fixed-target experiments are needed to lower these 
uncertainties. On the other hand, the interactions of primary cosmic rays in the atmosphere can be 
exploited to study the hadronic interaction models in a kinematic and energy region not accessible 
by human-made accelerators. Indeed, exploiting Auger data, we reach center-of-mass energies up 
to $\sqrt{s} \sim 400$ TeV, more than 30 times  those attainable at LHC and explore interactions in 
the very forward region of phase space on targets of $\left< A \right> \sim 14$.\\
The shower development depends on many different features of the hadronic interactions. 
In particular, by collecting the electromagnetic radiation emitted by the shower particles crossing the 
atmosphere and its depth of maximum development \xmax, we get information about the first interaction 
cross section \cite{pAir}. The measure of the muonic component at the ground is more sensitive to the 
details of the hadronic interactions along many steps of the cascade, like the 
multiplicity of the secondaries and the fraction of electromagnetic component with respect to the total signal. 
On the contrary, the intrinsic muon fluctuations mostly depend on the first interaction \cite{muf}.\\
Clear evidence for a deficit of the number of muons predicted by the models was reported by our 
Collaboration by exploiting inclined showers, where the electromagnetic component has been fully 
absorbed by the atmosphere \cite{mu1} and in the hadronic component \cite{mu2}. 
The study of the muon production depth \cite{mu3} and of the  time profiles of the signals recorded with 
SD \cite{delta} further proved the inability of models to describe all the components of the showers correctly. 

%%% FIG muUMD
\begin{figure}
  \includegraphics[width=.51\linewidth]{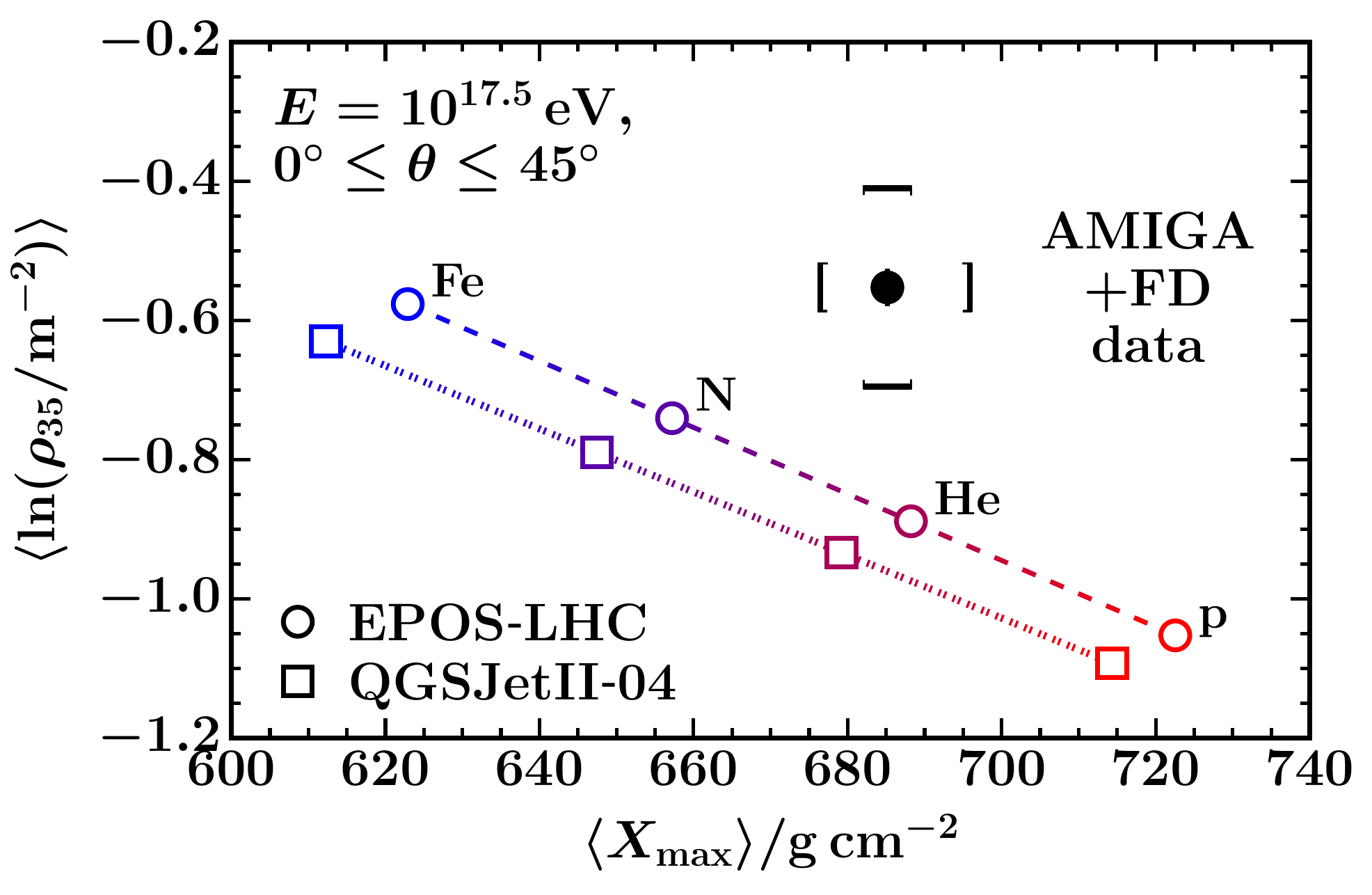}
  \includegraphics[width=.49\linewidth]{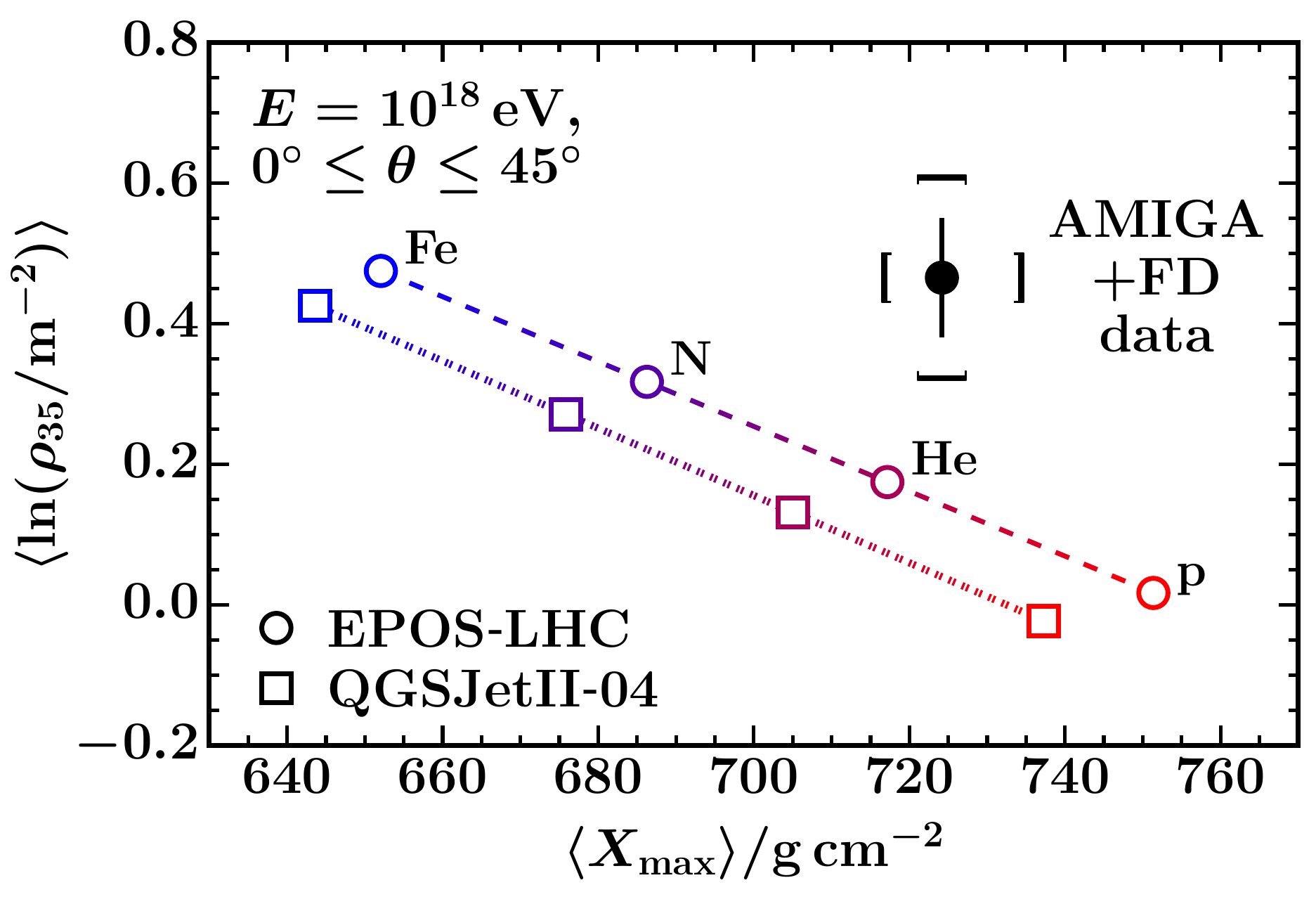}
\caption{Mean logarithmic muon density $\left< \textrm{ln} \rho_{35} \right>$vs \xmaxmed  expected by showers 
simulated at two different energies with EPOS-LHC and QGSJetII-04, in comparison to the
experimental results from UMD and FD.}
  \label{fig:muUMD}
\end{figure}
%%% FIG muFluct e Mean
\begin{figure}
  \includegraphics[width=.5\linewidth]{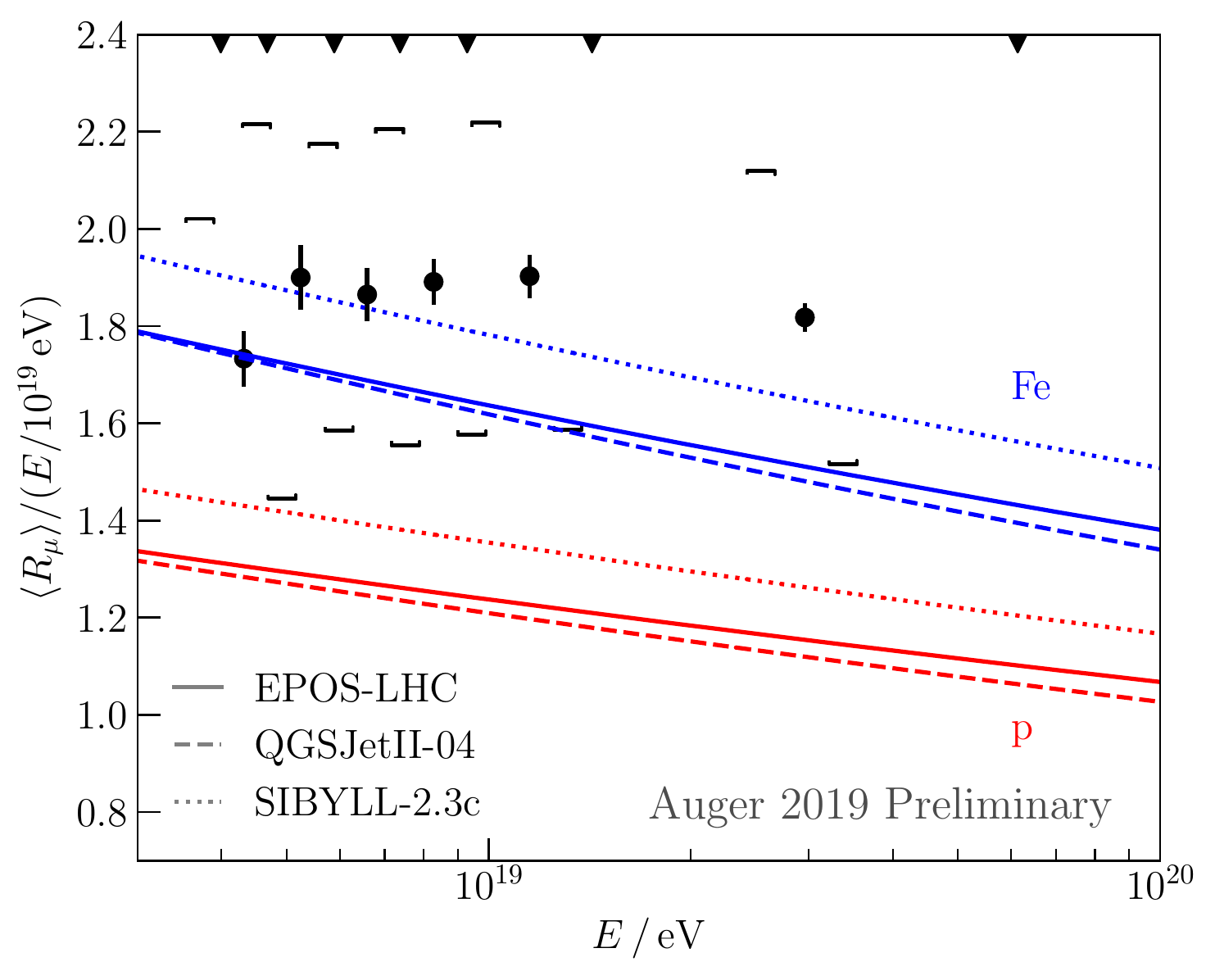}
  \includegraphics[width=.5\linewidth]{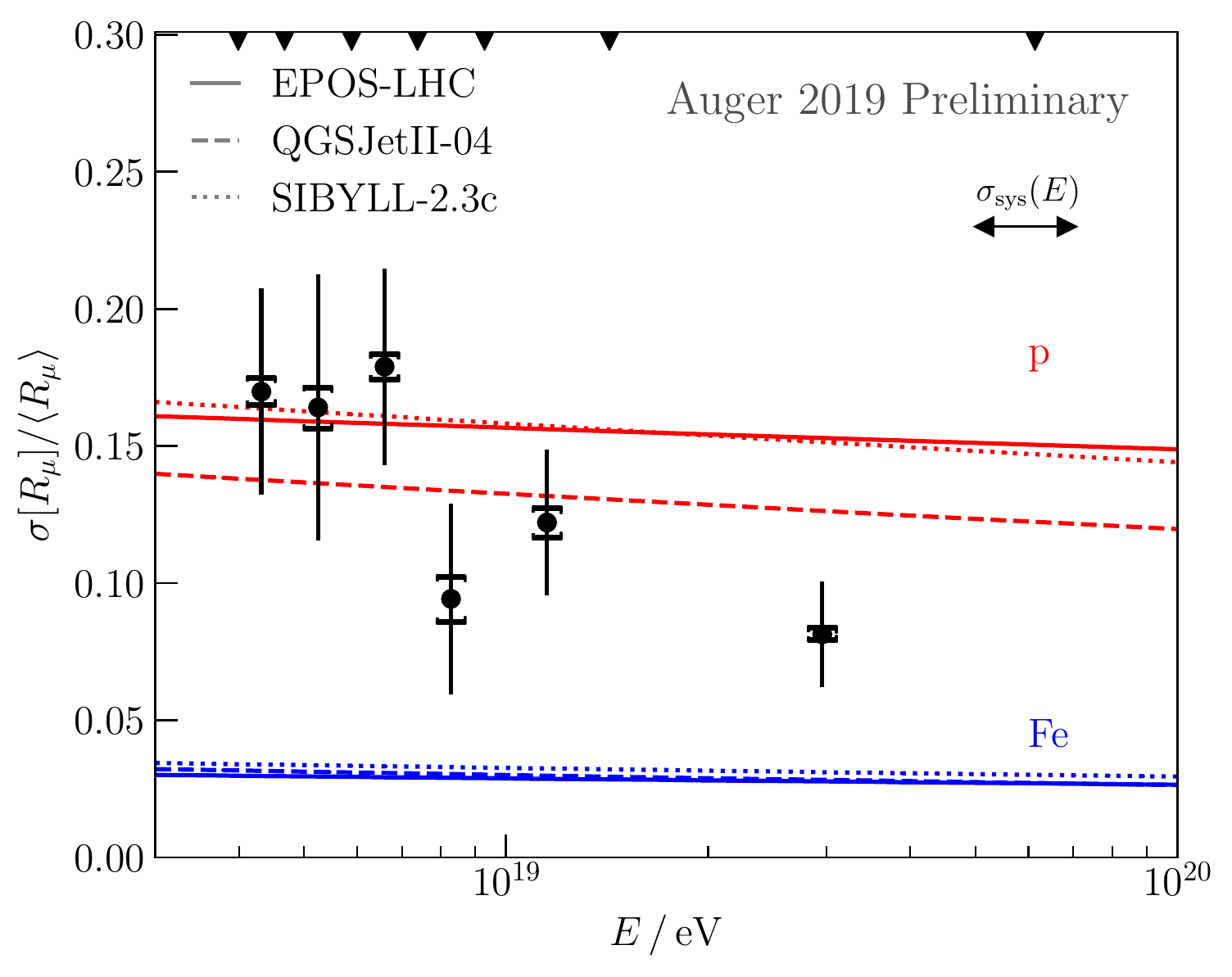}
\caption{Average relative muon number (left) and shower-to-shower fluctuations (right) in inclined events.}
  \label{fig:muflu}
\end{figure}
At this conference, we showed that the deficit of muons in the models is well visible also at lower energies, 
by directly measuring the muon content of EAS with an engineering array of underground muon detectors
(UMD) deployed in the infill area \cite{asp1}. 
In Fig.\ref{fig:muUMD}, the correlation of the muon densities measured by the UMD with the \xmaxmed 
measured by the FD is shown at two different energies ($10^{17.5}$ and $10^{18}$ eV). \\
The mean logarithmic muon densities are linearly dependent on the mean logarithmic primary mass, so 
that they can be related to the \xmaxmed in proton and iron simulations. 
The resulting predicted muon content is 38\% too low at both energies when EPOS-LHC is used, the mismatch
increasing to 50\% to 53\% at the two energies in the case of QGSJetII-04.

Moreover, we presented for the first time the measurement of the shower-to-shower fluctuations 
of the muonic component, at the same time updating the results on the mean muon number exploiting 
hybrid inclined showers \cite{asp2}. They are shown in Fig.\ref{fig:muflu}; note that  systematic errors 
(square brackets) are dominant for $\left< R_{\mu} \right>$, while for the muon fluctuations the uncertainties 
are mainly statistical (error bars).
The expectations for the relative fluctuations are compatible with the experimental results, while
models and data show a significant discrepancy in the average muon scale. 
The results thus suggest that models are describing reasonably well the distribution of energy going 
into the electromagnetic component after the first interaction, while the discrepancy in the overall muon 
number should be explained by resorting to changes of different hadronic interaction characteristics along 
all the shower stages.

Non-standard mechanisms, like  departures from relativity such as Lorentz invariance violation (LIV), 
can result in a perturbative modification of the particle dispersion relation  
$E_{i}^{2}=p_{i}^{2}+m_{i}^{2}+\sum_{n=0}^{\infty} \delta^{(n)}_{i}E_{i}^{n+2}$,
(where $i$=type of particle, $n$=approximation order). 
The UHECRs collected at the Pierre Auger Observatory, coming from extragalactic sources, 
represent the ideal candidates to search for these effects, as these are expected to be suppressed at 
low energies and for short travel distances.
At this conference, we presented searches for  Lorentz invariance violation effects in three different 
sectors \cite{pheno1}, by modifying the propagation of hadrons or photons or by considering the changes 
which would appear in the evolution of the showers in the atmosphere, and comparing the predictions on 
the energy spectrum and composition with those measured, in a combined fit procedure. 
In the first case, our data is better described with a low maximum rigidity 
at sources, which leads to a poor sensitivity to LIV effects. In the photon sector, the limits  of 
$\delta^{(1)}_{\gamma} \sim 10^{-40}$ eV$^{-1}$ and $\delta^{(2)}_{\gamma} \sim 10^{-60}$ eV$^{-1}$
obtained in our analysis are two and four orders of magnitude more restrictive than those in \cite{lang} for 
$n$=1 and $n$=2 respectively.

\section{Conclusion and future steps}
\label{sec:upgrade}

The rich harvest of results from the Pierre Auger Observatory presented at this conference is based 
on an unprecedented exposure and on a more and more careful understanding of our data and their 
systematic uncertainties.
The large-scale dipolar modulation above 8 EeV is confirmed, and  its amplitude is found growing with 
energy. The extension of the measurement of the amplitude and phase of the modulation to more than 
three orders of magnitude allowed us to confirm a Galactic origin for cosmic rays below the EeV, while 
the direction of the dipole strongly support the conclusion that they are extragalactic above few EeV.  
Blind searches for intermediate and small-scale anisotropies confirmed the existence of an excess in 
the direction of Centaurus A. A growing significance (now above 4.5$\sigma$) was found for the correlation 
among the directions of cosmic rays with $E>38$ EeV and the position of starburst galaxies in a catalog
including objects like NGC4945 and M83 in the CenA region, but also NGC253 near the Southern 
Galactic pole. More studies including the effects of magnetic fields and larger catalogs are required in the 
future to confirm these findings.

The measurement of the energy spectrum of cosmic rays covers a wide range of energies, from 
$\sim 10^{16.5}$ to above $10^{20}$ eV, and it is derived without assumptions on models or composition. 
Two inflection points are clearly visible corresponding to the second knee and to the ankle, a new 
feature around $10^{19}$ eV is now appearing thanks to the much reduced statistical errors.  
The region around the ankle appears to be dominated by a mixed composition, with a pure or too 
light one excluded with a significance of  $\sim 5 \sigma$ by our measurements based on hybrid events.
The best information on the primary mass relies on the evaluation of the \xmax distributions and their moments,
hinting to a composition which gets heavier above $2 \times 10^{18}$ eV, a conclusion confirmed by 
new measurements exploiting the 100\% duty cycle of the SD.  These results point to an 
origin for the confirmed steepening of the flux above $\sim 5 \times 10^{19}$ eV more related to the 
exhaustion of the sources rather than to propagation effects. Information about the mass composition 
in the suppression region is however still missing. 

New results on hadronic interaction models at ultrahigh energy come from the study of muons: a 
good description of the energy sharing after the first interaction is confirmed by the first-ever measurement 
of the evolution of the muon intrinsic fluctuations in extensive air showers. The deficit in the expected total 
number of muons is now confirmed also at energies down to $10^{17.5}$ eV thanks to the measurements 
performed with the UMD, and most probably due to effects related to the subsequent steps of the cascades 
in the atmosphere.  

The AugerPrime upgrade of the Pierre Auger Observatory \cite{prime} is being built to address the 
open questions raised by the current results.
The interpretation of the suppression in the flux by differentiating between a cut-off due to propagation 
effects and the maximum energy reached in the sources can provide  fundamental constraints on the 
sources of the ultrahigh energy cosmic rays and their properties.  To set the stage for future experiments 
and assess the feasibility of charged particle astronomy, it is mandatory to evaluate the possible existence 
of a fraction of protons at the highest energies. 
New information about hadronic interactions in a range of energy and in a kinematic region which is not 
accessible in human-made accelerators will allow us to break the degeneracy between primary mass 
and hadronic interactions and furthermore to search for non-standard mechanisms.  
These questions must be addressed not only by reaching larger exposures but more importantly by providing 
new information on the primary mass composition at ultra-high energies on an event-by-event basis.

Each station of the SD will be equipped by a 3.8 m$^2$ slab of plastic scintillator on top (the Surface Scintillator 
Detector, SSD \cite{prime4}), new SD electronics and an additional small photomultiplier inside the WCD 
for the extension of the dynamic range \cite{prime2,prime3}. Besides the scintillator, a radio antenna (RD) 
will be mounted above the station for the detection of the radio signals from the extensive air showers 
in the frequency range from 30 to 80 MHz\cite{prime1}.
Additionally, the UMD are going to be deployed beside each WCD in the infill 
region of the Observatory, to provide direct muon measurements \cite{found5}.  

The most direct technique to evaluate the muon component exploits the different responses of the 
WCD and SSD at station level, relating their total signals to the electromagnetic and muonic fluxes at 
the ground. Indeed, the ratio of the electromagnetic to muonic components of the showers is more than 
a factor of two higher in a thin scintillator with respect to the WCD over a very wide range of distances to 
the core. \\
This information can be obtained employing a matrix inversion method \cite{matrix}, although in this 
case the measured muon signals are subject to large fluctuations, being restricted to individual stations. 
A more general approach based on the properties of universality of air showers \cite{univ} or multivariate 
analyses would allow us to take advantage of all the information provided by AugerPrime, not only by correlating 
the detector signals at different distances to the core but also considering their temporal structure (arrival times 
of the particles at the stations). 
In this case, an event-by-event reconstruction will provide the energy,  the depth of maximum shower 
development in the atmosphere and the number of muons. 
Moreover, the implementation of the radio detector will allow us to collect a new different kind of hybrid 
events: the combination of the WCD and RD will be used to exploit the correlation of the electromagnetic 
energy (from radio) and the number of muons (from the WCD) for horizontal air showers. Indeed, the size of 
the radio footprint exceeds tens of km$^2$ for very inclined showers  ($\theta=60^{\circ}-84^{\circ}$) \cite{radio}, 
with dozens of stations involved.

Since 2016, 12 fully upgraded stations (the Engineering Array) are running in the field with new electronics, 
higher sampling, larger dynamic range. Furthermore, more than 350 upgraded stations (WCD+SSD) have 
already been deployed at the time of this conference. 
By exploiting the Engineering Array and a pre-production array of 80 upgraded stations running in the 
field with the old electronics, we could check the performances and calibration of the new detectors \cite{found3}.
The signals in the WCD are measured up to saturation ($\sim 650$ VEM) by the large PMTs. In the 
superposition region and above the large PMT saturation, they are derived exploiting the small PMT. 
The measured correlation among the signals of the WCD and SSD demonstrated the validity of the independent 
calibration procedures and showed that the required dynamic range in the upgraded SD is  nicely covered up 
to the highest particle densities ($\sim 20,000$ VEM). 
The differences of response of the two detectors were studied and confirmed by evaluating the dependency 
of the signal ratio with zenith, energy and distance to the core.

The prototypes of the RD are now installed in the field for in-situ evaluation of their performances, the 
starting of the mass production  is foreseen for the end of 2019.

The production and deployment of the AugerPrime detectors and electronics will be completed by 2020, 
for a data-taking planned up to 2025 and beyond.

\begin{multicols}{2}

\end{multicols}
\end{document}